\documentclass[a4paper,fleqn,usenatbib]{mnras}

\usepackage[T1]{fontenc}
\usepackage{ae,aecompl}

\usepackage{graphicx}	
\usepackage{amsmath}	
\usepackage{amssymb}	

\title[Multi-frequency study of PKS 2233$-$148]{Multi-frequency study of the gamma-ray flaring BL Lac object PKS~2233$-$148 in 2009--2012}

\author[A. B. Pushkarev et al.]{
A. B. Pushkarev,$^{1,2}$\thanks{E-mail: pushkarev.alexander@gmail.com (ABP)}
M. S. Butuzova,$^1$
Y. Y. Kovalev,$^{2,3,4}$
T.    Hovatta$^{5}$\\
$^1$Crimean Astrophysical Observatory, Nauchny 298409, Crimea, Russia\\
$^2$Astro Space Center of Lebedev Physical Institute, Profsoyuznaya 84/32, Moscow 117997, Russia\\
$^3$Moscow Institute of Physics and Technology, Dolgoprudny, Institutsky per. 9, Moscow region, 141700, Russia\\
$^4$Max-Planck-Institut f\"ur Radioastronomie, Auf dem H\"ugel 69, 53121 Bonn, Germany\\
$^5$Tuorla Observatory, Department of Physics and Astronomy, University of Turku, FI-20014 University of Turku, Finland\\\
}

\date{Accepted 2018 October 3. Received 2018 October 1; in original form 2018 August 17}
\pubyear{2018}

\begin{document}
\label{firstpage}
\pagerange{\pageref{firstpage}--\pageref{lastpage}}
\maketitle

\begin{abstract}
We study the jet physics of the BL Lac object PKS~2233$-$148 making use of synergy of observational 
data sets in the radio and $\gamma$-ray energy domains. The four-epoch multi-frequency (4--43~GHz) 
VLBA observations focused on the parsec-scale jet were triggered by a flare in $\gamma$-rays 
registered by the {\it Fermi}-LAT on April 23, 2010. We also used 15~GHz data from the 
OVRO 40-m telescope and MOJAVE VLBA monitoring programs. Jet shape of the source is found to be 
conical on scales probed by the VLBA observations setting a lower limit of about 0.1 on its unknown 
redshift. Nuclear opacity is dominated by synchrotron self-absorption, with a wavelength-dependent 
core shift $r_{\text{core\,[mas]}}\approx0.1\lambda_{[\text{cm}]}$ co-aligned with the innermost jet 
direction. The turnover frequency of the synchrotron spectrum of the VLBI core shifts towards lower 
frequencies as the flare propagates down the jet, and the speed of this propagation is significantly 
higher, about 1.2~mas~yr$^{-1}$, comparing to results from traditional kinematics based on tracking 
bright jet features. We have found indications that the $\gamma$-ray production zone in the source 
is located at large distances, 10--20~pc, from a central engine, and could be associated with the 
stationary jet features. These findings favour synchrotron self-Compton, possibly in a combination 
with external Compton scattering by infrared seed photons from a slow sheath of the jet, as a dominant 
high-energy emission mechanism of the source.
\end{abstract}

\begin{keywords}
galaxies: active -- 
galaxies: jets -- 
gamma-rays: galaxies --
BL Lacertae objects: individual: PKS~2233$-$148
\end{keywords}



\section{Introduction}
The location of the $\gamma$-ray production zone in active galactic nuclei (AGN) is still an open 
and actively debated question. Due to a limited angular resolution of the $\gamma$-ray telescopes 
it is impossible to directly locate the region responsible for the high-energy emission in AGN. 
A variaty of approaches has been considered to address this problem, and our current understanding
is that the regions of $\gamma$-ray production may be at different locations in different sources 
as evident from observations. One of the two main competing scenarios is based on the observed rapid 
$\gamma$-ray variability on time scales of a few hours and suggests that the high-energy emission 
from blazars is generated on sub-pc scales, near the central black hole \citep[e.g.,][]{Tavecchio10,Yan18}. 
Similarly, the observed strength and variability of the absorption of the $\gamma$-ray emission in 
the blazar 3C454.3 suggests the location of the $\gamma$-ray emitting zone within the broad-line 
region \citep{Bai09,Poutanen10}. The second scenario, in contrary, concludes that the dominant 
population of $\gamma$-ray photons is produced at larger, parsec scales, at distances up to 10--20~pc 
\citep{Marscher10,Agudo11,Schinzel12,Fuhrmann16,Karamanavis16}, and is based on a joint analysis 
of data in the $\gamma$-ray and radio bands. \cite{FM2} and \cite{Pushkarev10} showed that variability 
in $\gamma$-rays leads that of 15~GHz radio core on timescale of up to a few months.
In this paper we are concerned with one particular AGN, the BL Lac object 2233$-$148, which was
observed during and after the flare in $\gamma$-rays registered in April 2010 by the {\it Fermi}-LAT.

The structure of the paper is as follows: in Section~\ref{s:obs} we describe our and archival 
observational data and reduction schemes; in Section~\ref{s:results}, we discuss our results; 
and our main conclusions are summarized in Section~\ref{s:summary}. We use the term ``core''  
as the apparent origin of AGN jets that commonly appears as the brightest feature in VLBI 
images of blazars \citep[e.g.,][]{Lobanov_98}. The spectral index $\alpha$ is defined as 
$S_\nu\propto\nu^\alpha$, where $S_\nu$ is the observed flux density at frequency $\nu$. 
All position angles are given in degrees east of north. We adopt a cosmology with 
$\Omega_m=0.27$, $\Omega_\Lambda=0.73$ and $H_0=71$~km~s$^{-1}$~Mpc$^{-1}$ \citep{Komatsu09}.

\section{Observations and data processing}
\label{s:obs}

\subsection{Multi-epoch 4.6--43.2~GHz VLBA observations}
For the purposes of our study, we made use of data of the BL Lac object PKS 2233$-$148 
observed (code S2087D) with the Very Long Baseline Array (VLBA) of the National Radio
Astronomy Observatory (NRAO) during four sessions at epochs 2010-05-15, 2010-06-25, 
2010-08-01, and 2010-09-09. All ten VLBA antennas participated in each experiment.
The observations were performed in a full polarimetric mode simultaneously at C, X, 
U, K and Q frequency bands, which correspond to 6, 4, 2, 1.3, and 0.7~cm wavelengths, 
respectively (Table~\ref{t:freqs}). Each band was separated into four 8~MHz-wide 
intermediate frequency channels (IFs) having 16 spectral channels per IF. The signal 
was recorded with 2-bit sampling and total recording rate of 256~Mbps with analog base 
band converter. The data were correlated at the NRAO VLBA Operations Center in Socorro 
(New Mexico,USA) with averaging time of 2~sec. We split C and X bands into two sub-bands 
(each of 16~MHz width) centered at 4608.5, 5003.5~MHz and 8108.5, 8429.5~MHz. respectively,
and in the subsequent analysis the data were processed independently. U, K and Q bands 
were not split into sub-bands, resulting in 32~MHz band widths centered at 15365.5, 
23804.5 and 43217.5~MHz. On-source time at each epoch was, in total, about 45~min at C 
and X bands, 53~min at U and K bands, and 83~min at Q band split into 12 scans 
distributed over 8 hours. The scans are scheduled over a number of different hour angles 
to maximize the $(u,v)$ plane coverage. The increase of the on-source time with frequency 
was scheduled with the aim of obtaining comparable image sensitivity at all bands.

\begin{table}
\centering
\caption{Frequency setup for the S2087D VLBA experiment.}
\label{t:freqs}
\begin{tabular}{c r r r r}
\hline\noalign{\smallskip}
 Band & \multicolumn{4}{c}{Frequency channels} \\
      &     IF1 &     IF2 &     IF3 &     IF4 \\
      &   (MHz) &   (MHz) &   (MHz) &   (MHz) \\
\hline\noalign{\smallskip}
   C  &  4600.5 &  4608.5 &  4995.5 &  5003.5 \\
   X  &  8100.5 &  8108.5 &  8421.5 &  8429.5 \\
   U  & 15349.5 & 15357.5 & 15365.5 & 15373.5 \\
   K  & 23788.5 & 23796.5 & 23804.5 & 23812.5 \\
   Q  & 43201.5 & 43209.5 & 43217.5 & 43225.5 \\
\hline
\end{tabular}
\end{table}

The data reduction was performed with the NRAO Astronomical Image Processing System 
\citep[{\sc aips},][]{AIPS} following the standard procedure. The individual IFs for 
each frequency band were processed separately throughout the data reduction. The antenna 
gain curves and system temperatures measured during the sessions were used for a 
priori amplitude calibration. Global gain correction factors for each station for 
each IF were derived from the results of self-calibration. We applied the significant 
amplitude scale corrections listed in Table~\ref{t:gain_corr} by running the AIPS task 
{\sc clcor}. The phase corrections for station-based residual delays and delay rates were 
found and applied using the AIPS task {\sc fring} in two steps. First, the manual fringe 
fitting was run on a short interval on a bright quasar 3C454.3 (2251+158) to determine 
the relative instrumental phase and residual group delay for each individual IF. 
Secondly, the global fringe fitting was run by specifying a point-like source model 
and a signal-to-noise ratio cutoff of 5 to omit noisy solutions. The fringe-fit 
solution interval was chosen to be 10, 4, 2, 1.5, and 1 minute for C, X, U, K, and 
Q band, respectively. After fringe fitting, a complex bandpass calibration was made. 
The estimated accuracy of the VLBA amplitude calibration in the 5--15~GHz frequency 
range is of about 5\% and at 24--43~GHz of about 10\% 
\citep[see also][]{2cmPaperIV,Sokolovsky11}.

\begin{table}
\centering
\caption{Amplitude scale corrections for the S2087D VLBA experiment.
The full table is available online.}
\label{t:gain_corr}
\begin{tabular}{c c c c c}
\hline\noalign{\smallskip}
Antenna &          Band & Epoch &   IF & Correction \\
    (1) &           (2) &   (3) &  (4) &        (5) \\
\hline\noalign{\smallskip}
     BR &  K            &     1 & 1--2 &       0.88 \\
     BR &  K            &     1 & 3--4 &       0.85 \\
     BR &  K            & 2,3,4 & 1--4 &       0.80 \\
     FD &  U            &     1 & 1--4 &       1.09 \\
     FD &  Q            &     1 & 1--4 &       1.15 \\
\hline
\end{tabular}
\end{table}

{\sc clean}ing \citep{CLEAN}, phase and amplitude self-calibration \citep{Jennison58,Twiss60}, 
and hybrid imaging \citep{Readhead80,Schwab80,Cornwell81} were performed in the Caltech 
{\sc difmap} \citep{difmap} package. A point-source model was used as an initial model for the 
iterative procedure. Final maps were produced by applying a natural weighting of the 
visibility function. The spanned bandwidth of the IFs in each band is small ($<0.2$\% of
fractional bandwidth in all bands), thus no spectral correction technique was applied.

In this paper, we present results inferred from the total intensity images. 
The polarization calibration and results will be published in a separate paper.

\subsection{Multi-epoch 15.4~GHz MOJAVE observations}
We also made use of the data at 15.4~GHz from the MOJAVE (Monitoring of Jets in Active 
Galactic Nuclei With VLBA Experiments) program\footnote{\url{http://www.astro.purdue.edu/MOJAVE}}. 
The data were obtained at eight more epochs at 15.4~GHz: 2009-12-26, 2010-06-19, 2010-12-24, 
2011-09-12, 2012-05-24, 2012-07-12, 2012-12-10, and 2016-09-17. We used the fully calibrated 
publicly available data. For a more detailed discussion of the data reduction and imaging 
process schemes, see \cite{MOJAVE_XV}. The absolute flux density of the observations is 
accurate within 5\% \citep{MOJAVE_I,MOJAVE_VIII}.

\begin{figure*}
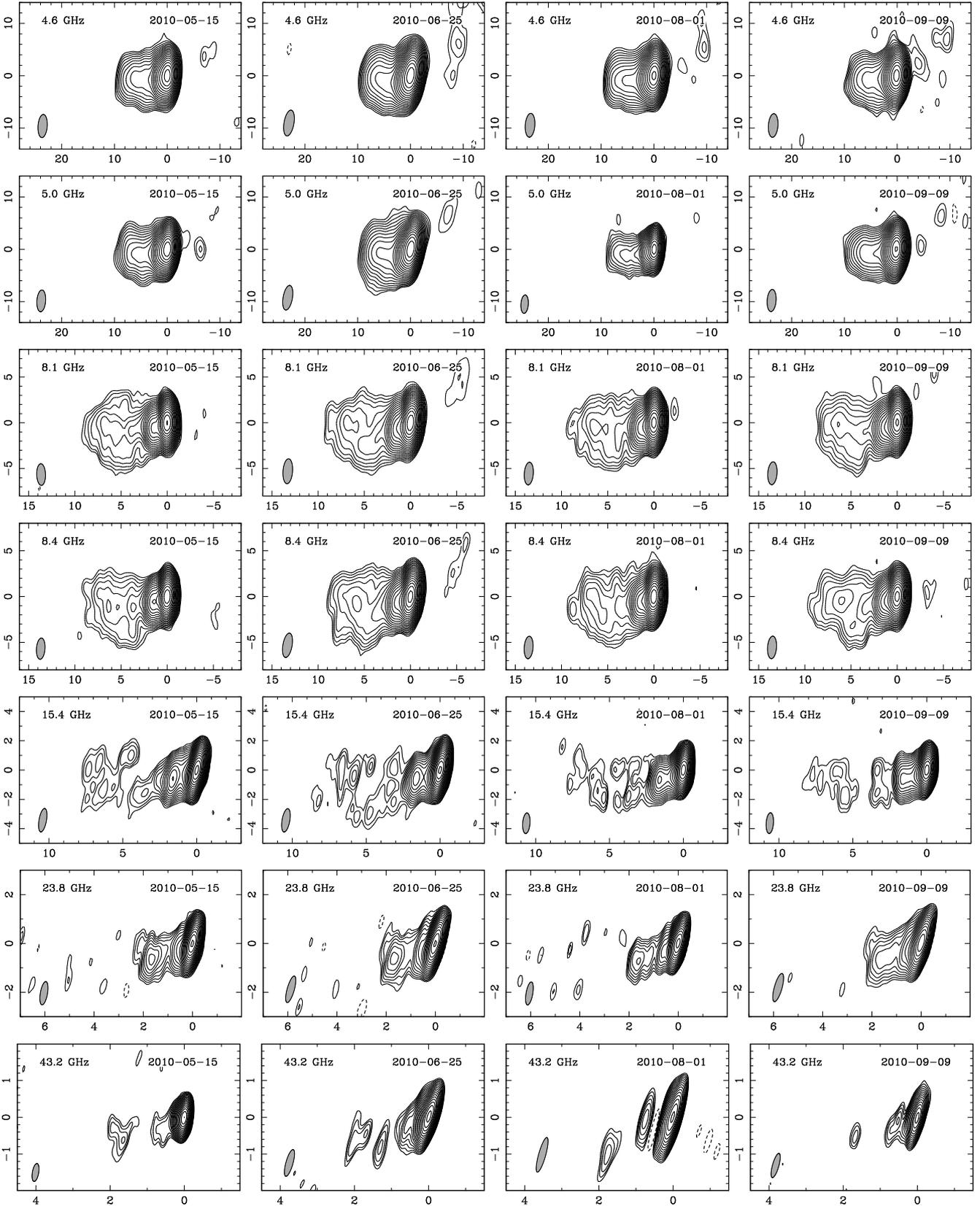

\centering
\includegraphics[width=2.92cm,angle=-90]{figs/01/J2236-1433_C1_2010_05_15_pus_map.ps}
\includegraphics[width=2.92cm,angle=-90]{figs/01/J2236-1433_C1_2010_06_25_pus_map.ps}
\includegraphics[width=2.92cm,angle=-90]{figs/01/J2236-1433_C1_2010_08_01_pus_map.ps}
\includegraphics[width=2.92cm,angle=-90]{figs/01/J2236-1433_C1_2010_09_09_pus_map.ps}\vspace{2mm}
\includegraphics[width=2.92cm,angle=-90]{figs/01/J2236-1433_C2_2010_05_15_pus_map.ps}
\includegraphics[width=2.92cm,angle=-90]{figs/01/J2236-1433_C2_2010_06_25_pus_map.ps}
\includegraphics[width=2.92cm,angle=-90]{figs/01/J2236-1433_C2_2010_08_01_pus_map.ps}
\includegraphics[width=2.92cm,angle=-90]{figs/01/J2236-1433_C2_2010_09_09_pus_map.ps}\vspace{2mm}
\includegraphics[width=2.92cm,angle=-90]{figs/01/J2236-1433_X1_2010_05_15_pus_map.ps}
\includegraphics[width=2.92cm,angle=-90]{figs/01/J2236-1433_X1_2010_06_25_pus_map.ps}
\includegraphics[width=2.92cm,angle=-90]{figs/01/J2236-1433_X1_2010_08_01_pus_map.ps}
\includegraphics[width=2.92cm,angle=-90]{figs/01/J2236-1433_X1_2010_09_09_pus_map.ps}\vspace{2mm}
\includegraphics[width=2.92cm,angle=-90]{figs/01/J2236-1433_X2_2010_05_15_pus_map.ps}
\includegraphics[width=2.92cm,angle=-90]{figs/01/J2236-1433_X2_2010_06_25_pus_map.ps}
\includegraphics[width=2.92cm,angle=-90]{figs/01/J2236-1433_X2_2010_08_01_pus_map.ps}
\includegraphics[width=2.92cm,angle=-90]{figs/01/J2236-1433_X2_2010_09_09_pus_map.ps}\vspace{2mm}
\includegraphics[width=2.92cm,angle=-90]{figs/01/J2236-1433_U0_2010_05_15_pus_map.ps}
\includegraphics[width=2.92cm,angle=-90]{figs/01/J2236-1433_U0_2010_06_25_pus_map.ps}
\includegraphics[width=2.92cm,angle=-90]{figs/01/J2236-1433_U0_2010_08_01_pus_map.ps}
\includegraphics[width=2.92cm,angle=-90]{figs/01/J2236-1433_U0_2010_09_09_pus_map.ps}\vspace{2mm}
\includegraphics[width=2.92cm,angle=-90]{figs/01/J2236-1433_K0_2010_05_15_pus_map.ps}
\includegraphics[width=2.92cm,angle=-90]{figs/01/J2236-1433_K0_2010_06_25_pus_map.ps}
\includegraphics[width=2.92cm,angle=-90]{figs/01/J2236-1433_K0_2010_08_01_pus_map.ps}
\includegraphics[width=2.92cm,angle=-90]{figs/01/J2236-1433_K0_2010_09_09_pus_map.ps}\vspace{2mm}
\includegraphics[width=2.92cm,angle=-90]{figs/01/J2236-1433_Q0_2010_05_15_pus_map.ps}
\includegraphics[width=2.92cm,angle=-90]{figs/01/J2236-1433_Q0_2010_06_25_pus_map.ps}
\includegraphics[width=2.92cm,angle=-90]{figs/01/J2236-1433_Q0_2010_08_01_pus_map.ps}
\includegraphics[width=2.92cm,angle=-90]{figs/01/J2236-1433_Q0_2010_09_09_pus_map.ps}
\caption{Naturally weighted total intensity contour maps of PKS~2233$-$148 at four epochs during 2010
         at 4.6, 5.0, 8.1, 8.4, 15.4, 23.8 and 43.2~GHz, with a cell size of 0.3, 0.3, 0.2, 0.2,
         0.1, 0.06, 0.03 mas per pixel, respectively. The x and y axes are given in mas of 
         relative right ascension and relative declination, respectively. The contours are plotted at 
         increasing powers of $\sqrt{2}$ starting from 4~rms level. The full width at half maximum 
         (FWHM) of the restoring beam is shown as a shaded ellipse in the lower left corner.
         Notice that the scales in the different images are different.
         The image parameters are listed in Table~\ref{t:maps}.}
\label{f:maps}
\end{figure*}

\subsection{15~GHz OVRO observations}
We also used public data\footnote{\url{http://www.astro.caltech.edu/ovroblazars}} of
PKS~2233$-$148 observations performed within the Owens Valley Radio Observatory 40-m
Telescope monitoring program \citep{Richards11}. Observations are done at 15~GHz in
a 3~GHz bandwidth since 2008-10-23 to 2018-02-05 with a typical time sampling of about
four days. Details of the data reduction and calibration are given in \cite{Richards11}.

\subsection{Gamma-ray {\it Fermi}-LAT data}
The $\gamma$-ray light curve was generated from data obtained with the LAT \citep{LAT}
onboard the {\it Fermi} $\gamma$-ray space telescope between 2008-08-09 and 2016-10-17.
In the analysis, we used the {\it Fermi} ScienceTools software
package\footnote{\url{http://fermi.gsfc.nasa.gov/ssc/data/analysis/documentation/Cicerone}}
version v10r0p5 and Pass 8 data. In generation of the light curve, we first selected 
all photons between 100 MeV and 300 GeV within a $10\degr$ region of interest (ROI) 
around the source. In the event selection and analysis we followed the recommendations 
for Pass 8 data, given by the LAT
team\footnote{\url{http://fermi.gsfc.nasa.gov/ssc/data/analysis/documentation/Pass8\_usage.html}}. 

The photon flux over each 7-day bin was calculated using the tool {\it gtlike} with 
instrument response function version P8R2\_SOURCE\_V6. The source model was generated 
using the external tool make3FGLxml.py version 01 by selecting all sources within $20\degr$
of the target in the 3FGL catalogue \citep{3FGL}, and including also the Galactic diffuse 
emission model version {\it  gll\_iem\_v06} and isotropic diffuse emission model version 
{\it iso\_source\_v06}. Based on 3FGL catalogue, the target was modeled with a log-parabola
spectrum, defined as $dN/dE=N_0(E/E_b)^{-(\alpha+\beta\log(E/E_b))}$. To account for low 
number of photons in each weekly bin and to reduce the number of free parameters in the fit, 
we froze the spectral parameters of the target and all other sources in the model to the 
values reported in 3FGL. For the target the 3FGL values are $\alpha=2.04$, $\beta=0.09$, 
and $E_b=581.68$. Additionally, if the source was beyond the $10\degr$ ROI or had a test 
statistic (TS) value \citep[e.g.,][]{Mattox96} less than five in 3FGL, we also froze the 
flux to the value reported in 3FGL. If the TS of the bin was less than four (corresponding 
to about 2$\sigma$) or if the number of predicted photons in that bin was
less than 10, we calculated a 95\% upper limit of the photon flux \citep{Abdo11}.

\section{Results}
\label{s:results}
\subsection{Parsec-scale jet structure}
Final naturally weighted VLBA maps of the source brightness distribution at the seven 
frequencies at each of the four observing epochs are presented in Fig.~\ref{f:maps}.
The source shows a typical parsec-scale AGN morphology of a bright compact core and 
one-sided jet, which propagates towards the east and is detected up to a distance of 
about 2~mas at 43~GHz and progressively farther, up to 8~mas at lower (4.6, 5.0~GHz) 
frequencies due to a steep spectrum of the jet emission (see more detailed discussion 
in Section~\ref{s:sp_ind}). At the frequency of 8~GHz and higher the outer jet regions 
are transversely resolved. The lower frequency images show the faint emission beyond 
the core, most probably caused by the uncompensated side-lobes due to low declination 
of the source. The images at 8~GHz are most sensitive, with a typical noise level of 
about 0.16~mJy beam$^{-1}$ and a dynamic range of the order of 3000, determined as a 
ratio of the peak flux density to the rms noise level. The noise level was calculated 
as the average of rms estimates in three corner quadrants of the image, each of 1/16 
of the map size. The forth quadrant, with a maximum rms was excluded being affected by 
the source structure.
In Table~\ref{t:maps}, we summarize the VLBA map parameters.

\begin{table*}
\caption{Summary of image parameters. Columns are as follows:
(1) epoch of observations,
(2) central observing frequency,
(3) I peak of image,
(4) rms noise level of image,
(5) theoretical thermal noise estimate,
(6) bottom I contour level,
(7) dynamic range of image,
(8) total flux density from map,
(9) FWHM major axis of restoring beam,
(10) FWHM minor axis of restoring beam,
(11) position angle of major axis of restoring beam.
The full table is available online.
}
\label{t:maps}
\begin{tabular}{c c c c c c c c c c r}
\hline\noalign{\smallskip}
Epoch &  Freq. & $I_\textrm{peak}$ &  $I_\textrm{rms}$ &   Thermal noise & $I_\textrm{base}$ &  DR &$S_\textrm{VLBA}$ & $B_\textrm{maj}$ & $B_\textrm{min}$ & $B_\textrm{PA}$ \\
      &  (GHz) &   (mJy bm$^{-1}$) &   (mJy bm$^{-1}$) & (mJy bm$^{-1}$) &   (mJy bm$^{-1}$) &     &            (mJy) &            (mas) &            (mas) &           (deg) \\
  (1) &    (2) &               (3) &               (4) &             (5) &               (6) & (7) &              (8) &              (9) &            (10)  &            (11) \\
\hline\noalign{\smallskip}
2010--05--15 &  4.608 &   335 &  0.19 &  0.11 &  0.76 &  1756 &   505 &  4.40 &  1.74 &  $-$2.0 \\
2010--06--25 &  4.608 &   408 &  0.15 &  0.11 &  0.61 &  2676 &   569 &  4.97 &  1.87 &  $-$7.6 \\
2010--08--01 &  4.608 &   382 &  0.17 &  0.11 &  0.68 &  2235 &   538 &  4.54 &  1.80 &  $-$3.5 \\
2010--09--09 &  4.608 &   357 &  0.21 &  0.11 &  0.83 &  1723 &   510 &  4.50 &  1.79 &  $-$2.0 \\
2010--05--15 &  5.003 &   350 &  0.18 &  0.15 &  0.71 &  1979 &   519 &  4.14 &  1.65 &  $-$3.3 \\
2010--06--25 &  5.003 &   413 &  0.15 &  0.15 &  0.60 &  2737 &   570 &  4.74 &  1.76 &  $-$8.1 \\
2010--08--01 &  5.003 &   371 &  0.30 &  0.15 &  1.19 &  1243 &   542 &  3.51 &  1.39 &  $-$2.2 \\
\hline
\end{tabular}
\end{table*}

Structure modeling of source brightness distribution was performed with the procedure 
{\it modelfit} in DIFMAP package by fitting several circular Gaussian components to 
the calibrated visibility data and minimizing $\chi^2$ in the spatial frequency 
plane. We used a minimum number of components (three at lower and four at higher 
frequencies) that after being convolved with the restoring beam, adequately reproduce 
the constructed source morphology. The obtained source models are listed in 
Table~\ref{t:models} and provide flux densities, positions, and sizes of the fitted 
components. All the positions are given with respect to the core component.

\begin{table*}
\caption{Source models. Columns are as follows:
(1) observation date,
(2) name of the component,
(3) flux density of the fitted Gaussian component,
(4) position offset from the core component,
(5) position angle of the component with respect to the core component,
(6) FWHM of the fitted circular Gaussian,
(7) SNR of the fitted Gaussian.
The full table is available online.
}
\label{t:models}
\begin{tabular}{c c c c c c r}
\hline\noalign{\smallskip}
      Date   & Comp. &     Flux density &        Distance &           P.A. &            Size &  SNR \\
             &       &             (Jy) &           (mas) &          (deg) &           (mas) &      \\
         (1) &   (2) &              (3) &             (4) &            (5) &             (6) &  (7) \\
\hline\noalign{\smallskip}
\multicolumn{7}{c}{4.6 GHz}\\
\hline
2010--05--15 &  Core &  $0.304\pm0.019$ & $0.000\phantom{\,\pm\,0.000}$ &         \ldots & $0.322\pm0.014$ &  535 \\
             &    J2 &  $0.127\pm0.012$ & $1.454\pm0.036$ & $113.7\pm 1.4$ & $1.009\pm0.070$ &  207 \\
             &    J1 &  $0.068\pm0.015$ & $5.049\pm0.515$ & $100.5\pm 5.8$ & $4.884\pm1.030$ &   23 \\
2010--06--25 &  Core &  $0.363\pm0.025$ & $0.000\phantom{\,\pm\,0.000}$ &         \ldots & $0.333\pm0.016$ &  428 \\
             &    J2 &  $0.127\pm0.015$ & $1.408\pm0.048$ & $114.2\pm 1.9$ & $1.092\pm0.094$ &  134 \\
             &    J1 &  $0.068\pm0.016$ & $5.012\pm0.557$ & $103.0\pm 6.3$ & $4.926\pm1.114$ &   20 \\
\hline
\end{tabular}
\end{table*}

\begin{figure}
\centering
\includegraphics[height=\columnwidth,angle=-90]{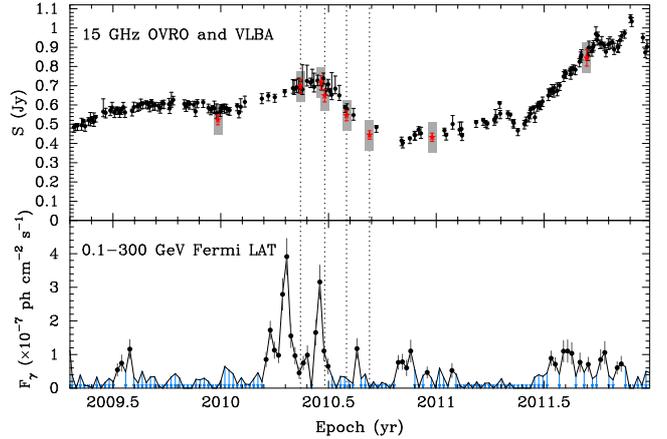}
\caption{OVRO 15~GHz flux density evolution (top panel). Grey rectangles together with red
         stars indicate VLBA total flux density at 15.4 GHz, observed within our campaign S2087D
         and the MOJAVE program. {\it Fermi} weekly-binned $\gamma$-ray light curve at 0.1--300~GeV
         (bottom panel). Upper limits are given by blue arrows. Dotted vertical lines indicate the
         epochs of the multi-frequency VLBA observations.
}
\label{f:lc}
\end{figure}

\subsection{Radio and $\gamma$-ray light curves}
In Fig.~\ref{f:lc}, we present light curves of PKS~2233$-$148 based on the {\it Fermi}-LAT and OVRO
monitoring data, complemented also by measurements of the MOJAVE program and our VLBA observations 
at 15~GHz. Prominent variability at high energies detected during April and June 2010 has triggered 
the four-epoch VLBA multi-frequency campaign. The values of the correlated VLBA total flux density 
are in good agreement with the single-dish OVRO flux density measurements, implying that there is 
almost no extended emission on kpc scales, as it was also previously concluded by \cite{Drinkwater97}. 
We performed a cross-correlation analysis of the light curves using the z-transformed discrete 
correlation function \citep{zDCF}, specifically developed for sparse, unevenly sampled light curves. 
The correlation between the radio and $\gamma$-ray light curves with and without upper limits is 
insignificant, suggesting that the $\gamma$-ray production region in the source might have a complex 
structure. We discuss it in more details in Section~\ref{s:gr_site}.

\begin{figure}
\centering
\includegraphics[height=\columnwidth,angle=-90]{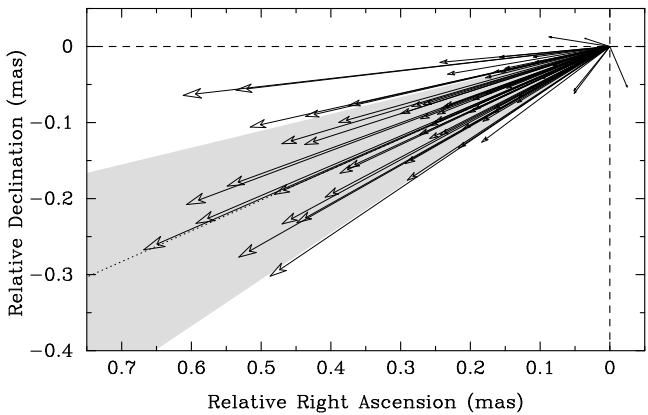}
 \caption{Core shift vectors measured in all frequency pairs.
 Typical error is 0.045~mas.
 Shaded grey area encompasses 68\% of the vectors deviating less than $10\degr$
 from the median jet direction shown by a dotted line.}
 \label{f:coreshifts}
\end{figure}

\subsection{Core shifts}
\label{s:coreshifts}

The VLBI core is believed to represent the apparent jet starting region, located at the distance 
$r_\text{core}$ to the central engine, at which its optical depth reaches $\tau_\nu\approx1$ at 
a given frequency. Thus, due to nuclear opacity, the absolute position of the radio core is 
frequency-dependent and varies as $r_\text{core}\propto\nu^{-1/k_\text{r}}$ \citep{BK79,Koenigl81}.
There is a growing observational evidence from recent multi-frequency studies of the core shift 
effect for $k_\text{r}\approx1$ \citep[e.g.,][]{Sullivan09,Fromm10,Sokolovsky11,Hada11_M87,Kravchenko16,Lisakov17}. 
This is consistent with the \cite{BK79} model of a synchrotron self-absorbed conical jet in 
equipartition between energy densities of the magnetic field and the radiating particles. 
Departures in $k_\text{r}$ from unity are also possible \citep{Plavin18_cs,Kutkin14} 
and can be caused by pressure and density gradients in the jet or by external absorption from the 
surrounding medium \citep{Lobanov_98,Kadler04}.

\begin{figure*}
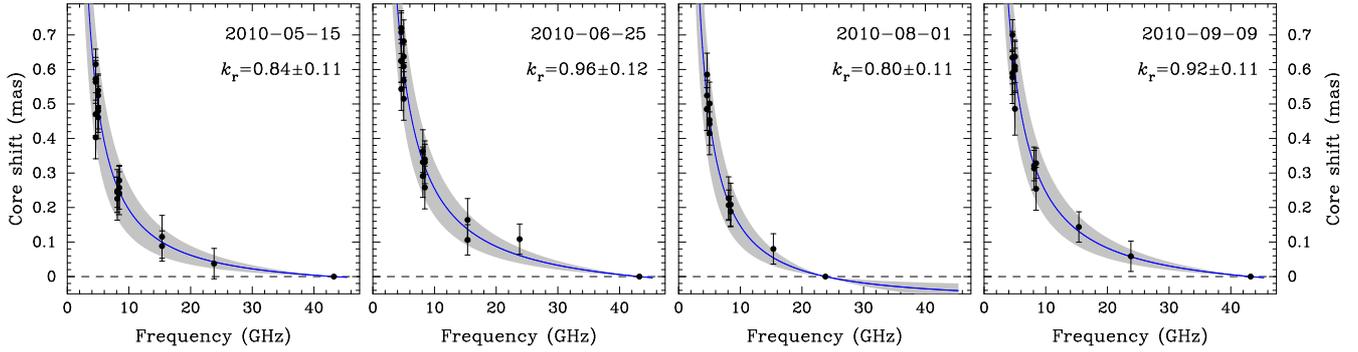

\centering
\includegraphics[width=4.56cm,angle=-90]{figs/04/coreshifts_da.ps}
\includegraphics[width=4.56cm,angle=-90]{figs/04/coreshifts_db.ps}
\includegraphics[width=4.56cm,angle=-90]{figs/04/coreshifts_dc.ps}
\includegraphics[width=4.56cm,angle=-90]{figs/04/coreshifts_dd.ps}
\caption{Frequency dependence of core shifts measured relative to the core position at
         43~GHz (23~GHz for the epoch 2010-08-01) for all observational multi-frequency epochs.
         Solid lines represent the best power-law fits.
         Shaded areas show $1\sigma$ confidence regions of the fit.}
\label{f:cs_vs_nu}
\end{figure*}

\begin{figure*}
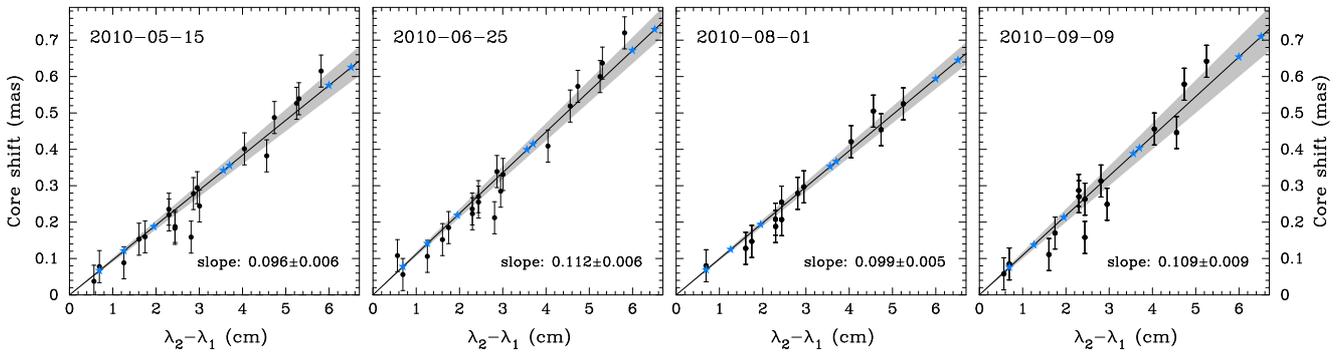

\centering
\includegraphics[width=4.56cm,angle=-90]{figs/05/cs_vs_delta_lambda_da.ps}
\includegraphics[width=4.56cm,angle=-90]{figs/05/cs_vs_delta_lambda_db.ps}
\includegraphics[width=4.56cm,angle=-90]{figs/05/cs_vs_delta_lambda_dc.ps}
\includegraphics[width=4.56cm,angle=-90]{figs/05/cs_vs_delta_lambda_dd.ps}
\caption{Core shifts as a function of difference of observing wavelengths. Solid lines
         represent the best linear fits. Shaded areas show $1\sigma$ confidence regions
         of the fit. Stars denote the expected core shifts from the true jet origin
         ($\lambda_1=0$) at wavelengths of our observations, 0.7, 1.3, 2.0, 3.6, 3.7, 
         6.0, 6.5~cm.}
\label{f:cs_vs_dl}
\end{figure*}

We calculated the core shift vector as
$\bmath{\Delta r}_{\text{core},\nu_1\nu_2} = \bmath{\Delta r}_{12} - (\bmath{r}_1 - \bmath{r}_2)$,
where $\bmath{\Delta r}_{12}$ is the displacement of the phase centers of the images at different
frequencies, and $\bmath{r}_1$, $\bmath{r}_2$ are VLBI core position offsets from the phase center.
To derive the image shift vector $\bmath{\Delta r}_{12}$, we used the fast normalized
cross-correlation algorithm \citep{Lewis95} to align the images to the same position on the sky,
selecting the jet regions of optically thin emission and assuming that their positions are
achromatic. Every pair of images was restored with the average beam size using a pixel size of
0.03~mas.

In Fig.~\ref{f:coreshifts}, we present a plot of 65 derived core shift vectors, where the head
of each vector represents the core position at lower frequency, while all core positions at higher
frequency are placed at the origin. The dotted line corresponds the median jet
direction of $\text{P.A.}=112\degr$. The core shift effect occurs predominantly along the jet
direction. In 68\% of cases, the core shift vectors deviate less than $10\degr$ from the median
jet position angle. This good alignment is achievable for a relatively straight jet, without
substantial curvature in the core region. Assuming that the core shift takes place along the jet 
and errors are random in direction, then the standard deviation of the transverse projections of 
the core shift vectors onto the jet direction yields the typical error of 45~$\mu$as. Thus, in 
90\% of cases the derived core shifts are significantly ($>2\sigma$) different from zero. In 
Table~\ref{t:coreshifts} we list the core shift measurements: (1) epoch of observations, (2) a pair 
of frequencies, (3) core shift magnitude, (4) core shift direction, (5) difference of observing 
wavelengths.

\begin{table}
\centering
\caption{Core shift vectors measured for the frequency pairs $\nu_1$ and $\nu_2$. 
         The full table is available online.}
\label{t:coreshifts}
\begin{tabular}{c c c r c}
\hline\noalign{\smallskip}
       Epoch & $\nu_1$ ~$\nu_2$ & $\Delta r_{\text{core},\nu_1\nu_2}$ & P.A.& $\lambda_2-\lambda_1$ \\
             &            (GHz) &                               (mas) &(deg)&                  (cm) \\
         (1) &              (2) &                                 (3) & (4) &                   (5) \\
\hline\noalign{\smallskip}
2010--05--15 & 43.2~   23.8 & 0.038 &   73 &  0.566 \\
2010--05--15 & 43.2~   15.4 & 0.088 &   82 &  1.258 \\
2010--05--15 & 43.2~ \,~8.4 & 0.278 &  106 &  2.865 \\
2010--05--15 & 43.2~ \,~8.1 & 0.244 &   95 &  3.006 \\
2010--05--15 & 43.2~ \,~5.0 & 0.539 &   96 &  5.302 \\
2010--05--15 & 43.2~ \,~4.6 & 0.615 &   96 &  5.816 \\
2010--05--15 & 23.8~   15.4 & 0.078 &  138 &  0.692 \\
2010--05--15 & 23.8~ \,~8.4 & 0.220 &  113 &  2.299 \\
\hline
\end{tabular}
\end{table}

We have studied the frequency dependence of the core shifts (Fig.~\ref{f:cs_vs_nu}) by 
fitting a function $\Delta r_\text{core} = b(\nu^{-1/k_\text{r}}-\nu_\text{max}^{-1/k_\text{r}})$,
where $b$ and $k_\text{r}$ are fitted parameters, and $\nu_\text{max}$ is fixed to the maximum 
frequency to which the core shifts were measured (43~GHz for the epochs 2010-05-15, 2010-06-25, 
and 2010-09-09; 23~GHz for the epoch 2010-08-01, at which we could not reliably measure core 
shift with respect to the core position at 43~GHz). The fitted $k_\text{r}$ values are smaller 
but close to one and not significantly differ from it. This can hold even during an outburst. 
As discussed in \cite{Plavin18_cs} a flare propagating down the jet disturbs only a 
limited portion of it, thus deviating $k_\text{r}$ from unity in a limited frequency range, 
significantly narrower compared to that of our observations. Therefore, for further analysis 
we use $k_\text{r}=1$. In this case, $r_\text{core}\propto\lambda$. Following the approach 
proposed in \cite{Voitsik18}, in Fig.~\ref{f:cs_vs_dl}, we plot the measured core shifts 
against difference in observing wavelengths (Table~\ref{t:coreshifts}) and fit the dependence 
by a straight line, from which one can also estimate an offset of the apparent jet base at a 
given wavelength from the true jet origin by setting $\lambda_1=0$.

\subsection{Jet shape}
\label{s:jet_shape}
\begin{figure}
\centering
\includegraphics[height=\columnwidth,angle=-90]{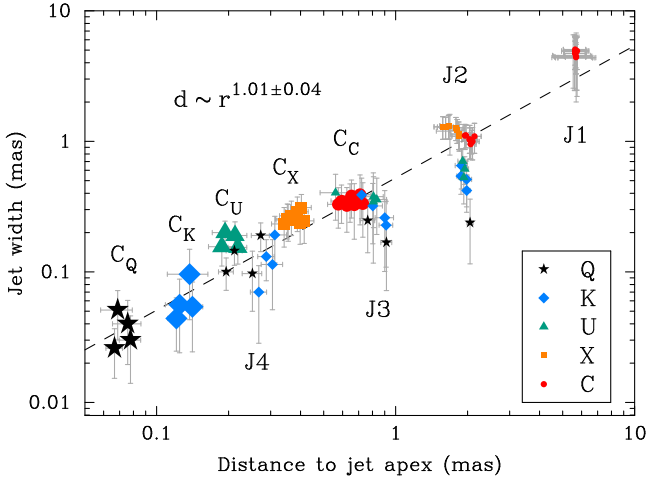}
\caption{Jet width versus distance to the jet vertex for 78 components from structure
         model fits at seven frequencies of four epochs. Cores are marked by larger
         symbols. Dashed line is the best fit from the least square method. The
         jet shape at scales probed by our multi-frequency VLBA observations is conical.}
\label{f:jet_shape}
\end{figure}

The core shift measurements allow us to perform a jet geometry analysis for the whole set of 
the fitted components including the cores. In Fig.~\ref{f:jet_shape}, we plot the transverse 
jet widths $d$ as the FWHM of the fitted Gaussian components at all four multi-frequency 
VLBA epochs (Table~\ref{t:models}) or the corresponding resolution limits \citep{2cmPaperIV} 
whichever is larger, as a function of their distance $r$ from the true jet 
base taking into account the core shift effect. The respective shifts 
$\Delta r_\text{core} = a(t)\lambda$ were added to the fitted core separations, where $a(t)$ 
is the fitted slope at a corresponding observational epoch (Fig.~\ref{f:cs_vs_dl}). From 
this analysis we excluded 16 weak components with $\text{SNR}<15$ to reduce the influence 
of low-SNR data points, though the whole set of 96 components yields qualitatively similar 
result. The BL~Lac object PKS~2233$-$148 shows a conical streamline, with 
$d\propto r^{1.01\pm0.04}$ at scales probed by the multi-frequency VLBA observations down
to 0.1~mas. This is consistent with $k_\text{r}\approx1$ derived from the core shift 
analysis. 

The apparent jet opening angle of the source is $38\degr\pm3\degr$, as reported by
\cite{MOJAVE_XIV}, who measured it from a stacked total intensity image at 15.4~GHz as a
result of combining VLBA maps from 11 epochs distributed over a time range of about 3~yr,
2009 through 2012. The wide opening angle suggests that the jet viewing angle is 
rather small, of the order of a few degrees.

\subsection{Redshift constraint from the jet geometry}
The optical spectrum of the source shows no prominent emission lines \citep{Drinkwater97} 
and its redshift is still unknown but the inferred conical jet shape indicates that this 
BL Lac object is not too close, and likely located at a redshift exceeding $\sim0.1$. 
Otherwise, our VLBA observations would be sensitive enough to reveal a jet geometry 
transition from parabolic to conical shape, as we detect this transition in a number of 
nearby ($z\lesssim0.1$) sources and explain it by a transition from magnetically to 
particle dominated regime in the outflows (Kovalev et al. 2018, in prep.). More distant 
AGNs ($z\gtrsim0.1$) typically show close to conical jet streamlines \citep{MOJAVE_XIV} 
since the scales probed by VLBI observations are beyond the shape transition region.

\subsection{Spectral index distribution}
\label{s:sp_ind}
\begin{figure}
\centering
\includegraphics[height=\columnwidth,angle=-90]{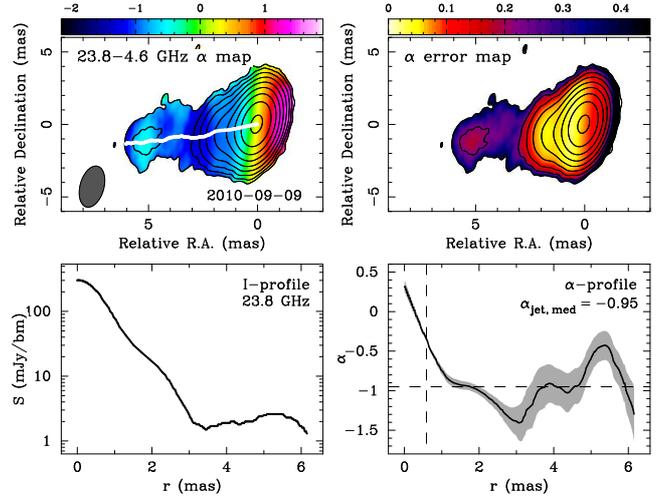}
\caption{ The distribution of the spectral index in PKS~2233$-$148 is shown in the top left 
          panel at the epoch 2010-09-09 calculated between 4.6 and 23.8~GHz; it is shown in 
          colour, with the 23.8~GHz total-intensity contours overlaid. The contours are plotted 
          at increasing powers of 2, starting from 0.35\% of the peak brightness of 
          303~mJy~beam$^{-1}$. White curve denotes the total intensity ridgeline. 
          The restoring beam is depicted, as a shaded ellipse in the lower left corner.
          Spectral index $1\sigma$ error map is shown in the top right panel. The bottom 
          left and right panels show total intensity and spectral index profiles along 
          the ridgeline, respectively. The vertical dashed line indicates the edge of the 
          convolved VLBI core along the inner jet direction, while the horizontal dashed 
          line represents the median jet spectral index. The grey area shows $1\sigma$ 
          errors on the spectral index.}
\label{f:alpha_map}
\end{figure}

The procedure of image registration by means of 2D cross correlation described in 
Section~\ref{s:coreshifts} allows to align the images at different frequencies and
accurately reconstruct a distribution of spectral index $\alpha$ over the source 
morphology. As an example, in Fig.~\ref{f:alpha_map} we present spectral index map of 
PKS~2233$-$148 calculated between 4.6 and 23.8~GHz at the epoch 2010-09-09 of our 
multi-frequency VLBA observations. It shows that the core is partially opaque, with 
spectral index about 0.3. while the outer jet regions are optically thin, with the 
median value of $\alpha_\text{jet}=-0.95$, which is typical for many other parsec-scale 
AGN jets \citep[e.g.,][]{RDV_paper,MOJAVE_XI}. The spectral index error map manifests
higher $\alpha$ accuracy in the innermost jet area and progressively larger 
uncertainties towards regions with lower brightness, where random errors dominate 
arising from the image noise. Systematic errors (from image alignment) dominate 
in the core area, especially behind it. Same result was obtained statistically for 
a large sample of sources in our earlier paper \citep[see Appendix B in][]{MOJAVE_XI}

\begin{figure}
\centering
\includegraphics[height=\columnwidth,angle=-90]{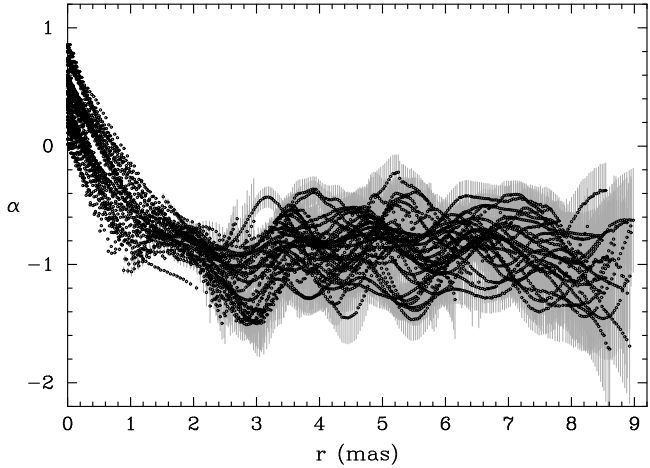}
\caption{
         Spectral index evolution along the ridgeline for $\alpha$-maps restored
         at all frequency pairs except the 43~GHz data.
         The grey bars show $1\sigma$ errors.
        }
\label{f:alpha_downstream}
\end{figure}

\begin{figure*}
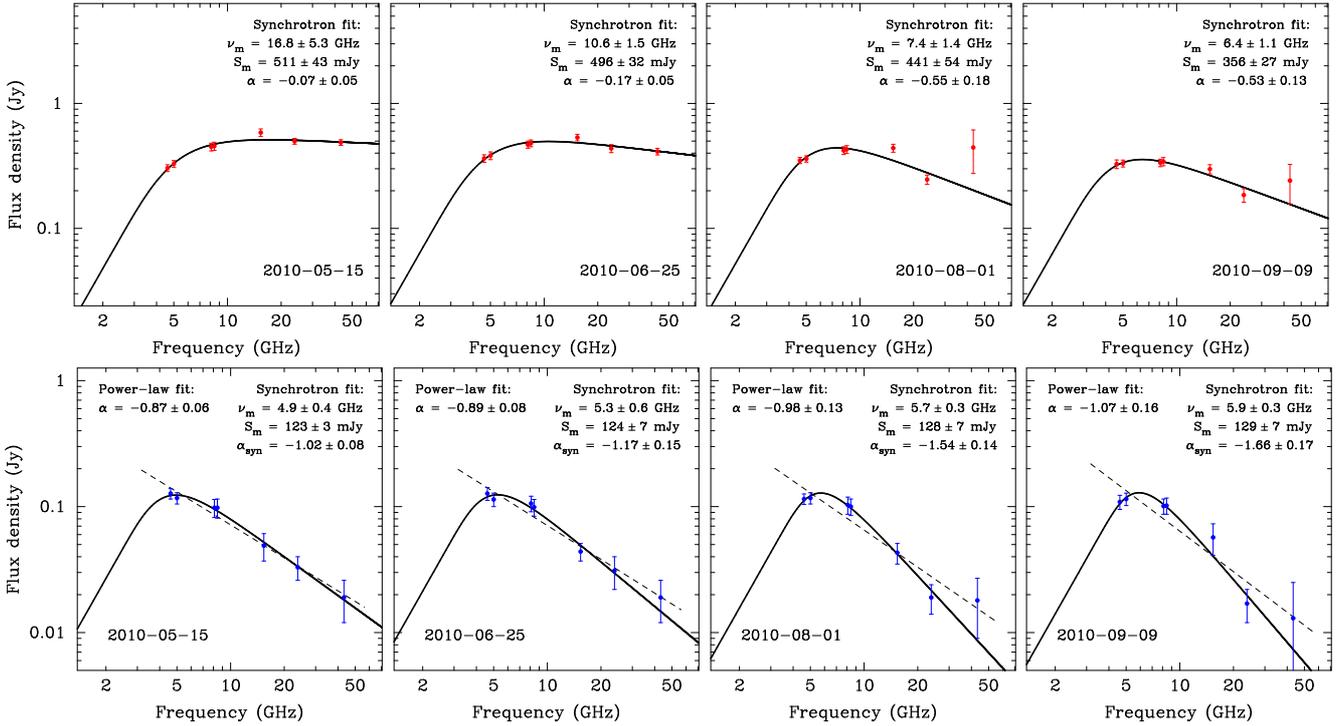

\centering
\includegraphics[width=4.7cm,angle=-90]{figs/09/J2236-1433_2010-05-15_spectrum_fit_core.ps}
\includegraphics[width=4.7cm,angle=-90]{figs/09/J2236-1433_2010-06-25_spectrum_fit_core.ps}
\includegraphics[width=4.7cm,angle=-90]{figs/09/J2236-1433_2010-08-01_spectrum_fit_core.ps}
\includegraphics[width=4.7cm,angle=-90]{figs/09/J2236-1433_2010-09-09_spectrum_fit_core.ps}\vspace{0.1cm}
\includegraphics[width=4.7cm,angle=-90]{figs/09/J2236-1433_2010-05-15_spectrum_fit_jet.ps}
\includegraphics[width=4.7cm,angle=-90]{figs/09/J2236-1433_2010-06-25_spectrum_fit_jet.ps}
\includegraphics[width=4.7cm,angle=-90]{figs/09/J2236-1433_2010-08-01_spectrum_fit_jet.ps}
\includegraphics[width=4.7cm,angle=-90]{figs/09/J2236-1433_2010-09-09_spectrum_fit_jet.ps}
\caption{Spectral fits to the core (top) and jet feature J2 (bottom) data. Solid lines represent 
         the spectra derived from the homogeneous synchrotron source model. Dashed lines show a 
         simple power-law model. Best fit parameters of models are shown on each plot.}
\label{f:spectra}
\end{figure*}

To analyze how spectral index changes
along the jet, we reconstructed the ridgeline of the outflow in total intensity using
a procedure described in \cite{MOJAVE_XIV}. As seen from Fig.~\ref{f:alpha_map} (bottom
right), the spectral index along the ridgeline slightly flattens in the jet knots
indicating reacceleration of emitting particles, while between them it is steeper.
Similar behaviour was found to be typical in AGN jets \citep{RDV_paper,MOJAVE_XI}.

Evolution of spectral index along the ridgeline for all the frequency pairs (except
Q-band data) taken at all four epochs is presented in Fig.~\ref{f:alpha_downstream}.
Beyond the core region, the spectral index deviates around the value of about $-1$.
The spectral index of optically thin synchrotron radiation parametrizes the energy 
spectrum of relativistic radiative particles. Assuming a power-law energy distribution 
$N(E)=N_0E^{-s}$, the power index $s=1-2\alpha$ has the mean value of $\sim3.0$.
The evolution of $\alpha_\text{jet}$ down the jet shows no effect of spectral aging 
(steepening downstream) owing to radiative losses of relativistic electrons 
\citep{Kardashev62}, which is often seen in AGN jets on parsec-scales \citep{RDV_paper}. 
In case of PKS~2233$-$148, the absence of this effect is likely caused by the dominance
of a few quasi-stationary jet features, which could be standing shocks that effectively 
accelerate the emitting particles.

\subsection{Synchrotron spectrum fitting and magnetic field estimates}
\label{s:sp_fits}
For spectral fitting of the core component (Table~\ref{t:models}), we use the standard 
spectrum of a homogeneous incoherent synchrotron source of relativistic plasma with a
 power-law energy distribution of the form $N(E)\propto E^{-s}$ \citep{Pacholczyk70}
\begin{equation}
S_\nu\propto\nu^{5/2}\left(1-\exp\left[-\left(\frac{\nu_1}{\nu}\right)^{5/2-\alpha}\right]\right)\,,
\end{equation}
where $\nu_1$ is the frequency at which the optical depth is $\tau=1$. The fitted spectra of
the core are presented in Fig.~\ref{f:spectra} (top). Best fit parameters, namely the optically 
thin spectral index $\alpha=(1-s)/2$, the peak flux density $S_\text{m}$ and the corresponding 
self-absorption turnover frequency $\nu_\text{m}$ are given for every spectrum. The core 
component shows a spectral turnover within the frequency range of our VLBA observations. 

With the fitted parameters $\nu_\text{m}$, $S_\text{m}$, and $\alpha$ of the synchrotron 
spectrum, we can estimate the magnetic field $B$ within the source adopting the 
standard synchrotron theory and assuming that the emission region is uniform and spherical. 
Then the commonly used expression of the component of the magnetic field perpendicular to 
the line of sight is (e.g. \cite{Marscher83}; see Appendix~\ref{a:mf_turnover} for more details)
\begin{equation}
B=10^{-5}\,b(\alpha)\,\theta_\text{m}^4\,\nu_\text{m}^5 S_\text{m}^{\prime\,-2}\,\left(\frac{\delta}{1+z}\right)\quad[G],
\label{eq:b_s}
\end{equation}
where $\delta$ is the Doppler factor, $z$ is the redshift, $\theta_\text{m}$ is the diameter of 
the spherical component at the turnover frequency, $S_\text{m}^\prime$ is the flux density at 
$\nu_\text{m}$ extrapolated from the straight-line optically thin slope, and $b(\alpha)$ 
(Fig.~\ref{f:b_alpha}) is a function of spectral index $\alpha$, optical depth $\tau_\text{m}$ 
at $\nu_\text{m}$, physical constants and a conversion factor, which allows one to express 
$\nu_\text{m}$ in GHz, angular size $\theta_\text{m}$ in mas, and $S^\prime_m$ in Jy. To derive 
$\theta_\text{m}$ we applied logarithmic interpolation between the measured component sizes at 
different frequencies (Table~\ref{t:models}) and multiplied the result by a correction factor of 
1.8 \citep{Marscher87} to take into account that the emission feature is assumed to have a spherical 
shape, while performing model fitting we measure the FWHM of the circular Gaussian components. 
The flux density $S_\text{m}^\prime$ can be calculated using the following relation:
\begin{equation}
S_\text{m}^\prime=S_\text{m}e^{\tau_\text{m}}\,,
\end{equation}
where the optical depth at the turnover can be derived numerically from the equation
$\exp(\tau_\text{m})=1+\tau_\text{m}(1-2\alpha/5)$
or approximated as \citep{Tuerler99}
\begin{equation}
\tau=\frac{3}{2}\left(\sqrt{1-\frac{16\alpha}{15}}-1\right)\,.
\end{equation}

The mean value of the magnetic field inferred from Eq.~(\ref{eq:b_s}) for the apparent core 
at the turnover frequency is $(0.03\pm0.01)\delta/(1+z)$~G. 

The only jet component detected at all the frequencies is J2 (see Fig.~\ref{f:jet_shape} and 
Table~\ref{t:models}). It is located at a distance of about 2~mas from the core at 43~GHz. 
The spectra of this jet feature J2 in addition to a synchrotron fit was also fitted by a simple 
power-law (Fig.~\ref{f:spectra}, bottom). The spectrum of J2 is steep, with a spectral index 
gradually decreasing from about $-1$ to $-1.6$, while the turnover frequency slightly 
increases from $4.9\pm0.4$~GHz to $5.9\pm0.3$~GHz.

\subsection{Evolution of the turnover frequency and source kinematics}

The self-absorption turnover frequency derived from the core spectra (Fig.~\ref{f:spectra}, 
top) gradually decreases from $16.8\pm5.3$~GHz on May 15, 2010 to $6.4\pm1.1$~GHz on September 
9, 2010 following inverse proportionality to time ($\nu_\text{m}\propto t^{-1}$), as predicted
by a model of a conical jet with constant plasma speed \citep{Blandford90}. We interpret these 
changes as a direct observational evidence of a flare propagation downstream. Due to synchrotron 
opacity in the nuclear region, the flare developing along the jet becomes detectable at progressively 
larger distances from the true jet origin corresponding to the VLBI core locations $r_\text{core}$ 
at longer wavelengths. After the disturbance crossed the apparent core ($\tau\approx1$ zone) at a 
given frequency, its flux density starts to decrease resulting in steepening the core spectrum 
(Fig.~\ref{f:spectra}, top) due to energy losses to synchrotron radiation or Compton scattering 
\citep{Kardashev62,MG85}. The core offset from the jet apex can be calculated as 
$r_\text{core}=r_\text{flare}=a(t)\lambda_\text{m}(t)$, where the parameter $a(t)$ was derived 
from the core shift analysis for each epoch of the multi-frequency VLBA observations 
(Fig.~\ref{f:cs_vs_dl}). 

\begin{figure}
\centering
\includegraphics[height=\columnwidth,angle=-90]{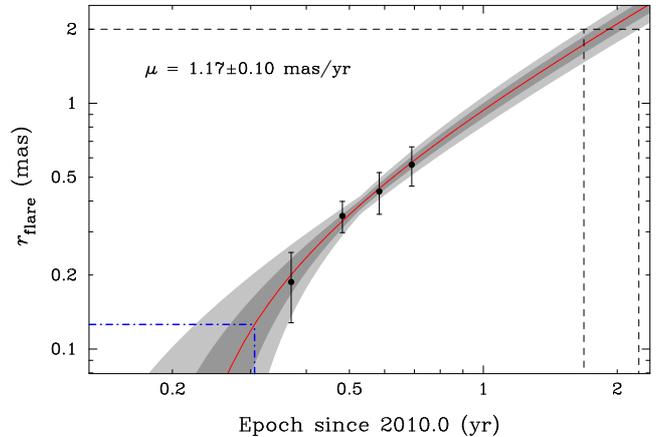}
\caption{Flare propagation down the jet. The measurements denote distances from the jet origin to
         the VLBI core at the frequency of maximum emission (Fig.~\ref{f:spectra}, top) at the
         four observing VLBA epochs. Solid line represents the best linear fit. Dark and light
         shaded area show $1\sigma$ and $2\sigma$ confidence regions of the fit, respectively.
         Dot-dashed lines indicate the epoch 2010.31 of the $\gamma$-ray flare and the
         corresponding distance 0.12~mas of the $\gamma$-ray emission zone from the central
         engine. Dashed lines indicate the expected epoch range (2012) when the flare reaches
         the jet component J2 at 2~mas separation from the true jet base assuming constant 
         flare propagation speed.
      }
\label{f:flare_prop}
\end{figure}

In Fig.~\ref{f:flare_prop}, we show $r_\text{flare}(\lambda_\text{m})$ as a function of time. 
The slope of the weighted linear fit is $1.17\pm0.10$~mas~yr$^{-1}$. The derived proper motion 
of the flare propagation is significantly higher than that inferred from kinematics analysis 
based on tracing bright jet components. The two jet knots, J2 and J3, of the source studied within 
the MOJAVE program at 15~GHz are slow pattern features, with the angular speed $-71\pm35$~$\mu$as 
(apparent inward motion) and $45\pm41$~$\mu$as \citep{MOJAVE_XIII}, respectively. These components 
are quasi-stationary and can be standing recollimation shocks observed in sources with 
super-magnetosonic jets \citep[e.g.,][]{Asada12,Cohen14} and also obtained in numerical 
two-dimensional relativistic (magneto)-hydrodynamic simulations \citep{Mizuno15,Fromm15,Fuentes18}.

\begin{figure}
\centering
\includegraphics[width=\columnwidth,angle=0,trim = 0.5cm 0cm 1.3cm 1.5cm,clip]{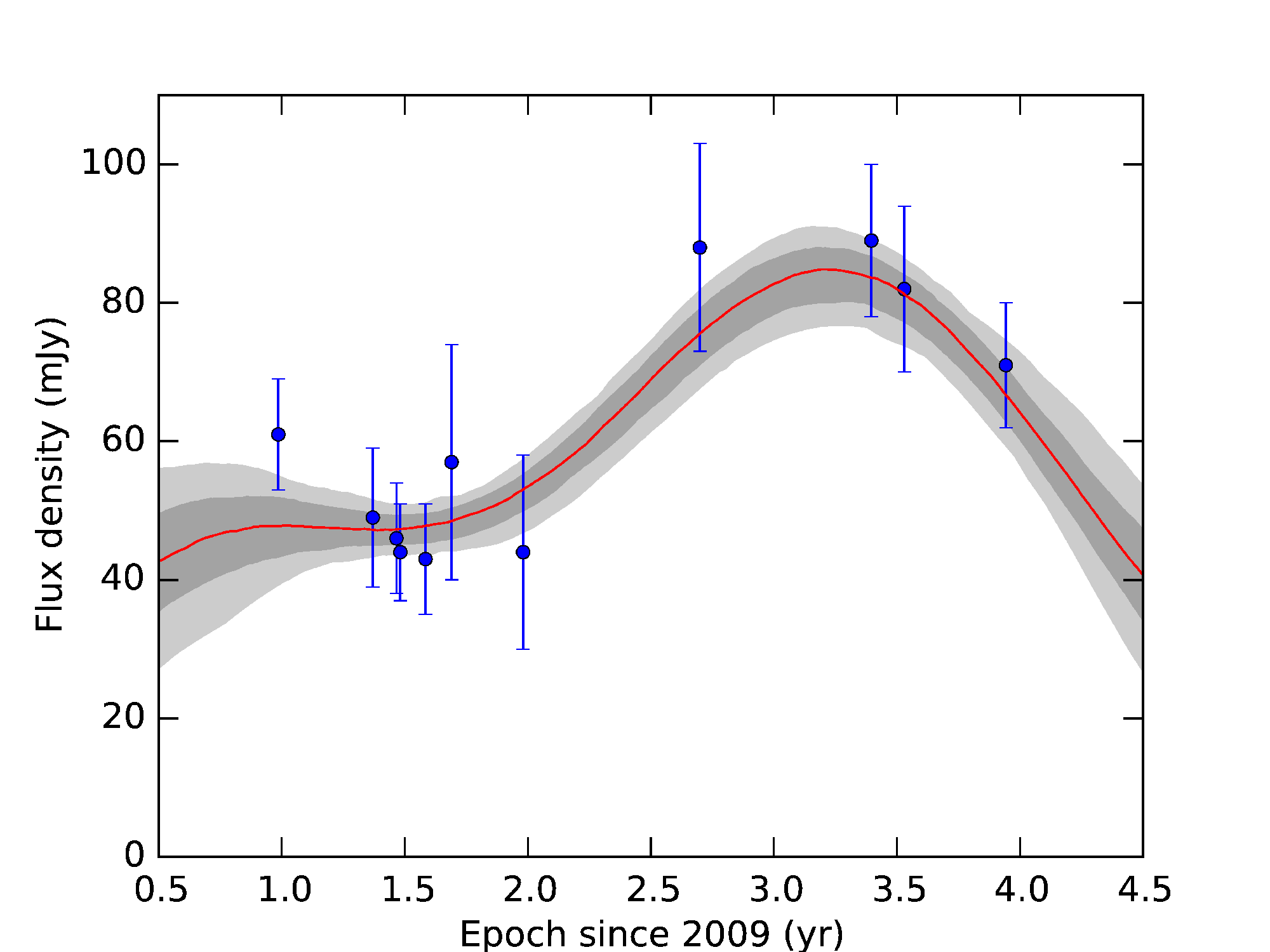}
\caption{Flux density at 15~GHz of the jet component J2 located at 2~mas from the jet origin.
         The significant increase of the flux density in 2012 is likely due to the flare
         which originated in 2010 and reached this jet region in 2012, as expected from 
         the assumed constant flare propagation speed. Error bars represent $1\sigma$
         uncertainties of individual measurements. Smooth solid curve shows the Gaussian
         process fit. Dark and light filled areas correspond to 68\% and 95\% confidence
         intervals of the fit, respectively.}
\label{f:comp_flare}
\end{figure}

Since the jet geometry is found to be conical at scales probed by the VLBA observations 
(Sec.~\ref{s:jet_shape}), we assume that the regime of constant flow speed holds at least 
up to 6~mas from the jet apex. Then the expected epoch for the flare to reach the 
quasi-stationary jet feature J2 at a distance of about 2~mas from the true jet base is 
$\sim$ 2012.0 (Fig.~\ref{f:flare_prop}), at which the turnover frequency is expected to 
decrease down to about 1.5~GHz. This scenario is supported by a flux density evolution of 
the component at 15~GHz (Fig.~\ref{f:comp_flare}). The component shows an increase of the 
flux density by a factor of about 2 during a period of time since $\sim2011.5$ through 
$\sim2012.5$. The epoch of the peak around 2012.0 was established by fitting the data with 
Gaussian process regression performed with the PyMC3 python module for Bayesian modeling, 
for which we used exponential quadratic covariance function. Note that moving jet components 
behave in a completely different manner. Typically, their brightness rapidly fades due to 
energy losses and adiabatic expansion \citep[e.g.,][]{RDV_paper,Kravchenko16,MOJAVE_XIII}.

It is therefore possible that the flare propagation rate represents the bulk flow speed. Taking 
into account the lower limit on redshift ($z>0.49$) derived from spectroscopy of the absorption 
lines formed by the intervening gas \citep{Sbarufatti06}, the proper motion of the disturbance 
$\mu=1.17\pm 0.10$~mas~yr$^{-1}$ corresponds to apparent speed $\beta_\text{app}>34\pm2\,c$. It 
is much faster than a typical apparent speed $\approx4\,c$ derived from kinematics analysis for 
a sample of 42 BL Lacs and also significantly higher than $\beta_\text{app}^\text{max}=21\,c$ 
detected in the high-redshift ($z=1.07$) BL Lac object 1514+197 \citep{MOJAVE_XIII}. 
Similarly, analyzing multi-frequency time delays of the flares and measuring the core shifts 
in the blazars 3C~454.3 \citep{Kutkin14} and 0235+164 \citep{Kutkin18} it was found that this 
approach yields the source jet speed, which is by a factor of a few higher than the estimates 
based on kinematic analysis.

Now, we assume that the jet of PKS~2233$-$148 in the core region is in equipartition between 
the particle and magnetic field energy density ($k_\text{r}=1$), has a spectral index 
$-0.5$, and is viewed at the critical angle $\theta\simeq\Gamma^{-1}$. Then the magnetic field 
in Gauss at 1~pc of actual distance from the jet apex can be estimated using the following 
relation \citep{MOJAVE_IX}
\begin{equation}
B_1\simeq0.04\,\Omega_{r\nu}^{3/4}\,(1+z)^{1/2}(1+\beta_\text{app}^2)^{1/8}\,,
\end{equation}
where $\Omega_{r\nu}$ is the core shift measure defined in \cite{Lobanov_98} as:
\begin{equation}
\Omega_{r\nu}=4.85\cdot10^{-9}\,\frac{\Delta r_\mathrm{core,\,\nu_1\nu_2}\,D_L}{(1+z)^2}\cdot\frac{\nu_1\nu_2}{\nu_2-\nu_1}\ \mathrm{pc \cdot GHz},
\end{equation}
where $\Delta r_\mathrm{core,\,\nu_1\nu_2}$ is the core shift in milliarcseconds, $D_L$ is 
the luminosity distance in parsecs, and $\beta_\text{app}=1.58\times10^{-8}D_L\mu/(1+z)$. The 
magnetic field strength at the apparent VLBI core at a given frequency can be calculated as 
$B_\text{core}=B_1r_\text{core}^{-1}$, where the absolute distance in parsecs of the core from 
the true jet base $r_\text{core}$ is given by 
\cite{Lobanov_98}
\begin{equation}
r_\text{core}(\nu)=\frac{\Omega_{r\nu}}{\nu\sin\theta}\approx\frac{\Omega_{r\nu}(1+\beta^2_\text{app})^{1/2}}{\nu}\,,
\end{equation}
where $\nu$ is the observed frequency in GHz. 

\begin{figure}
\centering
\includegraphics[height=\columnwidth,angle=-90]{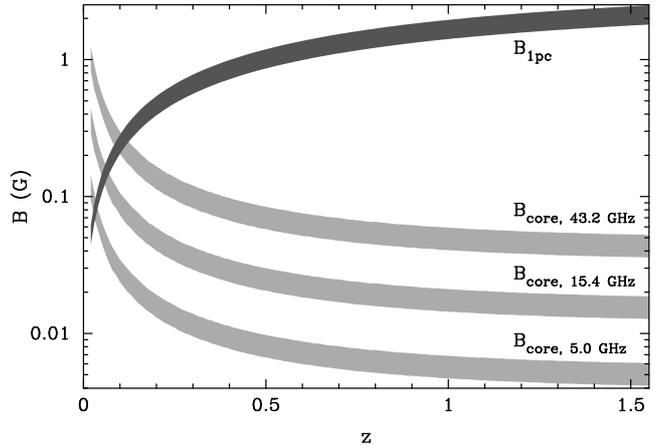}
\caption{Magnetic field constraints obtained from the core shift measurements and the flare
         propagation speed. Dark grey area show estimates of the magnetic field at 1~pc
         distance from the jet origin, while light grey stripes represent the magnetic
         field at the core at 43,2, 15,4 and 5.0~GHz.}
\label{f:B_cs}
\end{figure}

In Fig.~\ref{f:B_cs}, we plot the derived $B_1$ and $B_\text{core}$ for the 43, 15, and 5~GHz cores 
as functions of redshift. Assuming $z>0.5$ the magnetic field at a distance of 1~pc from the central 
engine is of the order of 1~G. We note that the estimates of $B_\text{core}$ at 15~GHz derived from 
the core shift analysis are comparable (lower by a factor of a few) to those inferred from the 
synchrotron spectrum fits (Sec.~\ref{s:sp_fits}), if a source Doppler factor is moderate 
($\delta\lesssim5$), as often observed in BL Lacertae objects \citep{Hovatta09,Liodakis17}.

\subsection{Location of the $\gamma$-ray emission region and the source of seed photons}
\label{s:gr_site}

To estimate the location of the $\gamma$-ray emission zone in PKS~2233$-$148 we extrapolated the 
$r_\text{flare}$ dependence (Fig.~\ref{f:flare_prop}) back to the epoch of the $\gamma$-ray flare, 
2010.31 (Fig.~\ref{f:lc}). This yields an angular separation of $0.12\pm0.03$~mas from the true jet 
base, which corresponds to the VLBI core position at 24~GHz (Fig.~\ref{f:jet_shape}). On a linear 
scale, this separation is $>0.7\pm0.2$~pc in projection, which exceeds a de-projected separation of 
$8\pm2$~pc if we assume a jet viewing angle of $5\degr$. Considering the second peak of the 
$\gamma$-ray data at epoch 2010.46 (June 17, 2010), we inferred distance of about 0.3~mas from the 
jet apex. This distance corresponds to the innermost jet feature J4 detected at 24 and 43~GHz setting 
an absolute distance of about 20~pc for another possible location of the $\gamma$-ray emission site. 
These assessments favour a scenario for the $\gamma$-ray production zone to be located at large 
distances from a central energy generator (beyond the broad-line region or torus) and likely 
associated with one or more standing shocks in a relativistic outflow of PKS~2233$-$148, as hinted 
by a complex structure of the major high-energy flares in the source. Similar conclusion regarding 
remotness of the $\gamma$-ray emission region in blazars on scales of parsecs from the central black 
hole was also made from other arguments in a number of recent single-source studies of 1510$-$089 
\citep{Marscher10}, OJ~287 \citep{Agudo11}, 3C~345 \citep{Schinzel12}, CTA~102 \citep{Casadio15}, 
1502+106 \citep{Karamanavis16}, BL Lacertae \citep{Wehrle16}, and also statistical results from 
F-GAMMA project \citep{Fuhrmann16}. At the same time, the arguments based on short-scale variability 
and breaks in GeV spectra discussed in Introduction indicate that the high-energy production site 
is in the immediate vicinity of the black hole.

\begin{figure}
\centering
\includegraphics[height=\columnwidth,angle=-90]{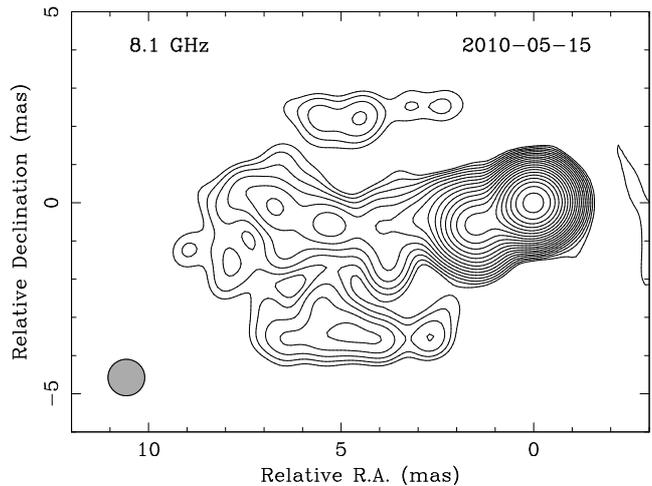}
\caption{Total intensity map of PKS 2233$-$148 at 8.1~GHz at the epoch 2010-05-15 from Fig.~\ref{f:maps}
         but convolved with a circular beam of a size of the minor axis of the original image. Emission
         at the jet edges indicate a presence of a distinct boundary layer.
        }
\label{f:map_sheath}
\end{figure}

Another noticeable feature of the pc-scale morphology of the source is a presence of a sheath around
the jet, the indications of which are seen at 8 and 15~GHz maps (Fig.~\ref{f:maps}) that provide a 
combination of a high angular resolution and sensitivity. To better visualize the sheath emission
we convolved the 8.1~GHz image at the epoch 2010-05-15 with a circular beam setting its FWHM to
that of the minor axis of the original map (Fig.~\ref{f:map_sheath}). The fact that this sheath is 
slower than a central spine might mean that it acts as a source of additional seed photons for the 
$\gamma$-ray radiation \citep[e.g.,][]{Marscher10,Aleksic14}. Thus, the high-energy emission of the 
source can be formed through (i) synchrotron self-Compton mechanism acting in its relativistic 
outflow and upscattering of low-energy synchrotron seed photons and (ii) external Compton scattering 
due to a photon field in the sheath.

While detailed modeling of the spectral energy distribution (SED) is beyond the scope of this paper, 
based on the publicly available non-simultaneous data in the SSDC SED builder 
tool\footnote{\url{https://tools.ssdc.asi.it}}, it seems that when the source is in a high state in 
$\gamma$-rays, the luminosity of the inverse Compton peak is higher than that of the synchrotron peak, 
indicating that an additional external photon field is indeed needed. However, this should be verified 
with simultaneous data in all bands, taken at both low and high activity states of the source.

We can also speculate that if instead of the epoch of the $\gamma$-ray peak we consider the epoch 
when the flare starts rising (a few weeks before the peak), then the $\gamma$-ray emission site 
could be in the immediate vicinity of the central machine. But this scenario is vulnerable as a plasma 
cloud moving fast down the jet leaves the seed photon area rapidly, while the flare is still reaching 
its maximum.

\section{Summary}
\label{s:summary}

We performed a radio and $\gamma$-ray joint study of the BL Lacertae object PKS~2233$-$148, using multiwavelength 
data in the period of 2009--2012. The 4.6--43.2~GHz VLBA observations reveal the core dominated, one-sided 
and relatively straight jet morphology of the source extending up to 8~mas in a position angle $112\degr$. 
Analyzing jet widths derived from the structure model fits we have established that the outflow has a conical 
shape. This sets a lower limit of about 0.1 on the source unknown redshift. 

We have measured the frequency-dependent shift vectors of the apparent core position using a method based on 
results from (i) structure model fitting and (ii) image alignment achieved by implementing a two-dimensional 
cross-correlation technique on the optically thin jet regions. The magnitude of the core shifts ranges from 
0.04 to 0.7~mas, with a typical uncertainty of 45~$\mu$as. The directions of the shift vectors are predominantly 
aligned with the median jet position angle, deviating from it by $\lesssim10\degr$ in 68\% of cases. The derived 
core shifts show a frequency dependence $\propto\nu^{-1/k_\text{r}}$, with $k_r\approx1$ indicating that nuclear 
opacity is dominated by synchrotron self-absorption, and physical conditions in the jet on scales probed by the 
VLBA observations are close to equipartition. We did not find an evidence for significant changes in $k_\text{r}$ 
between the observing epochs covering a time scale of four months, during which a flare was developing down the 
jet. It suggests that the transverse size of the disturbance area is significantly smaller than the jet part 
constrained by the magnitude of the core shift effect within a frequency range 5--43~GHz. The VLBI core position 
$r_\text{core}$, as a function of wavelength follows an $r_\text{mas}^\text{core}\approx0.1\lambda_\text{cm}$ 
dependence. The magnetic field at a distance of 1~pc from the jet apex derived from the core shift measurements
is about 1~G.

We present a method of independent assessment of jet kinematics based on core shift measurements and evolution
of synchrotron spectrum of the VLBI core. The turnover frequency of the core spectrum linearly shifts towards 
lower frequencies with time, as the flare originated in April 2010 in $\gamma$-rays propagates down the jet. 
The speed of this propagation is about 1.2~mas~yr$^{-1}$ and likely represents the bulk flow speed. It is 
much higher comparing to results from traditional kinematics based on tracking bright jet features, 
0.045~mas~yr$^{-1}$ \citep{MOJAVE_XIII}. 

We have found indications that the $\gamma$-ray production zone in the source is located at large distances, 
10--20~pc, from a central engine, and can be associated with the stationary radio-emitting jet features 
observed with VLBI. This favours synchrotron self-Compton scattering as a dominant high-energy radiation 
mechanism in the relativistic jet of the source. Direct observational evidence for a boundary layer around 
the jet suggests that the sheath might be an additional source of seed photons for external Compton scattering
acting in the source.

\section*{Acknowledgements}
We would like to thank the anonymous referee as well as E.~Ros for useful comments 
and suggestions. The VLBA data 
processing and core shift analysis were supported by the Russian Science Foundation 
grant 16-12-10481. The radio/$\gamma$-ray joint analysis was supported by the 
Academy of Finland projects 296010 and 318431. T.H. acknowledges support from the 
Turku Collegium of Science and Medicine. This research has made use of data from 
the MOJAVE data base, which is maintained by the MOJAVE team \citep{MOJAVE_XV}. 
The MOJAVE project was supported by NASA-\textit{Fermi} GI grants NNX08AV67G, 
NNX12A087G, and NNX15AU76G. This work made use of the Swinburne University of 
Technology software correlator \citep{DiFX}, developed as part of the Australian 
Major National Research Facilities Programme and operated under licence. This 
research has made use of data from the OVRO 40-m monitoring program 
\citep{Richards11} which is supported in part by NASA grants NNX08AW31G, 
NNX11A043G, and NNX14AQ89G and NSF grants AST-0808050 and AST-1109911. The 
National Radio Astronomy Observatory is a facility of the National Science 
Foundation operated under cooperative agreement by Associated Universities, Inc.




\bibliographystyle{mnras}
\bibliography{pushkarev}


\newpage\clearpage


\appendix


\section{Magnetic field from synchrotron self-absorption}
\label{a:mf_turnover}

Interpretation of a radio spectrum with the low-frequency turnover caused by
synchrotron self-absorption, and determination of physical parameters within 
this assumption date back to the 1960s \citep[see, for example, one of the 
pioneering works][]{Slysh63}. In particular, magnetic field associated with 
a source of synchrotron emission can be inferred. However, the approximate 
values of the numerical coefficient in the formula that are a function of a 
spectral index $\alpha$ ($S_\nu \propto \nu^\alpha$) of the optically thin 
part of a synchrotron spectrum are tabulated for a limited number of $\alpha$ 
values ranging from $-0.25$ to $-$1.0 \citep{Marscher83}. The relation for  
this coefficient was previously discussed by \cite{Gould79}. In this Appendix, 
we derive a formula for this coefficient that can be computed precisely.

Following \cite{Pacholczyk70}, intensity of emission in a case of synchrotron 
self-absorption is
\begin{equation}
S_\nu=F \left( \nu_1 \right) J \left( \frac{\nu}{\nu_1},s \right)\,,
\label{eq: FSSA}
\end{equation}
where $s=1-2\alpha$ is the power-law index of energy distribution $N \left( E \right)=N_0 E^{-s}$
of emitting electrons, and
\begin{equation}
J \left( \frac{\nu}{\nu_1},s \right)=\left(\frac{\nu}{\nu_1} \right)^{5/2} 
\left\{ 1- \exp \left[-\left(\frac{\nu}{\nu_1} \right)^{-\left(s+4 \right)/2} \right] \right\}\,,
\label{eq: J}
\end{equation}
where $\nu_1$ is the frequency at which the optical depth $\tau=1$. Source 
function $F \left( \nu_1 \right)$ for a spherical, uniform emitting region with observed angular 
size $\theta=2R(1+z)^2/D$ at the luminosity distance $D$ is
\begin{multline}
F \left(\nu_1 \right)=\frac{ \left( \pi \theta^2 R /3 \right) \varepsilon_{\nu_1} 
\left[\delta/\left( 1+z\right) \right]^{2-\alpha}}{2 \kappa _{\nu_1} R 
\left[\delta/\left(1+z \right) \right]^{\left( 3 - 2 \alpha\right)/2}} =\\
=\frac{\pi}{6} \theta^2 \frac{\varepsilon_{\nu_1}}{\kappa_{\nu_1}} \left( 
\frac{\delta}{1+z}\right)^{1/2},
\label{eq:Sor}
\end{multline}
where
\begin{equation}
\varepsilon_{\nu_1}=c_5\left(s \right)N_0 B_{\bot}^{\left( s+1\right)/2} \left(\frac{\nu_1}{2 c_1} \right)^{\left( 1-s \right)/2}
\label{eq:emis}
\end{equation}
and
\begin{equation}
\kappa_{\nu_1}=c_6\left(s \right) N_0 B_{\bot}^{\left(s+2 \right)/2} \left(\frac{\nu_1}{2 c_1} \right)^{-\left( s+4 \right)/2}
\label{eq:abs}
\end{equation}
are the emission and absorption coefficients, respectively, $B_\bot$ is 
the component of the magnetic field perpendicular to the line of sight. 
Constants and functions are:
\begin{equation}
c_1=\frac{3e}{4 \pi m^3 c^5}\,, \qquad
\label{eq:c1}
c_3=\frac{\sqrt{3} e^3}{4 \pi m c^2}\,,
\end{equation}
\begin{equation}
c_5=\frac{1}{4} c_3 \frac{s+7/3}{s+1}\,\, \Gamma \left(\frac{3 s-1}{12} \right) \Gamma \left(\frac{3 s+7}{12} \right)\,,
\label{eq:c5}
\end{equation}
\begin{equation}
c_6=\frac{1}{32} \left(\frac{c}{c_1} \right)^2 c_3 \left(s+\frac{10}{3} \right) \Gamma \left(\frac{3 s+2}{12} \right) \Gamma \left(\frac{3 s+10}{12} \right),
\label{eq:c6}
\end{equation}
where $e$ and $m$ are the charge and mass of electron, respectively, $c$ is 
the speed of light in vacuum, and $\Gamma$ is the Euler gamma function.

Substituting Eqs.~(\ref{eq: J})-(\ref{eq:c6}) into (\ref{eq: FSSA}) we obtain 
\begin{equation}
S_\nu=\frac{\pi}{6}\,\frac{c_5 \left( s \right)}{c_6 \left( s \right)} \left(2 c_1 
\right)^{-5/2} \theta^2 B_\bot^{-1/2} \nu^{5/2} \left(\frac{\delta}{1+z} \right)^{1/2} 
\left[ 1- \text{e}^{-\tau_\nu} \right],
\label{eq:Fnu}
\end{equation}
where $\tau_\nu=\left(\nu/\nu_1 \right)^{-\left(s+4 \right)/2}$.
Then the flux density at the turnover frequency $\nu_{\text{m}}$ extrapolated from 
the straight-line slope of the optically thin part of synchrotron spectrum is

\begin{multline}
S_\text{m}^\prime=\left.S_\nu \right|_{\nu=\nu_m} \text{e}^{\tau_\text{m}}= \\
=\left[\frac{\pi}{6} \frac{c_5 \left( s \right)}{c_6 \left( s \right)} \left( 2 c_1 
\right)^{-5/2} (e^{\tau_\text{m}}-1) \right] \theta^2 B_\bot^{-1/2} \nu_\text{m}^{5/2} \left(\frac{\delta}{1+z} \right)^{1/2}\,.
\label{eq:S_extr}
\end{multline}

\begin{figure}
\centering
\includegraphics[height=0.7\columnwidth,angle=-90]{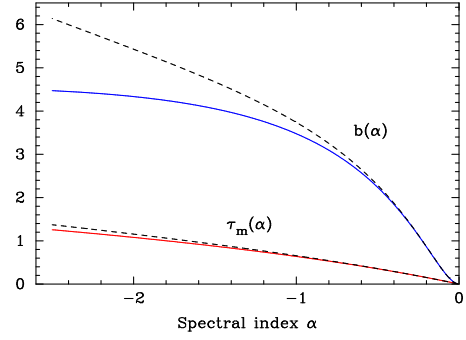}
 \caption{Coefficient $b(\alpha)$ and optical depth $\tau_\text{m}(\alpha)$ as a function of
          spectral index $\alpha$. Dashed lines represent approximate solutions derived
          by using Eq.~\ref{eq:Tuerler}.}
 \label{f:b_alpha}
\end{figure}

The optical depth $\tau_\text{m}$ at $\nu_\text{m}$ is found from equation  
\begin{equation}
\text{e}^{\tau_\text{m}}=1+\left(1-\frac{2\alpha}{5}\right)\tau_\text{m}\,.
\label{eq:tau}
\end{equation}
By developing the exponential in Eq.~(\ref{eq:tau}) to the third order, 
\citet{Tuerler99} obtained the approximate solution
\begin{equation}
\tau_\text{m}=\frac{3}{2} \left(\sqrt{1-\frac{16 \alpha}{15}}-1 \right)\,,
\label{eq:Tuerler}
\end{equation}
which is accurate enough, deviating less than 5\% from the exact numerical 
solution 
for $\alpha \gtrsim -1.55$ (Fig.~\ref{f:b_alpha}).

Solving Eq.~(\ref{eq:S_extr}) for the normal component of magnetic field, we have
\begin{equation}
B_\bot=10^{-5}\,b \left(s \right)\, \theta^4\, S_\text{m}^{\prime\,-2}\, \nu_\text{m}^5 \left(\frac{\delta}{1+z} \right)\,,
\label{eq:B}
\end{equation}
where the source size $\theta$ is in milliarcseconds, flux density in Jy, 
the turnover frequency $\nu_{m}$ in GHz, and the magnetic field in G.
\begin{equation}
b \left( s \right) = 5.5\cdot10^{62}\left[ \frac{\pi}{6} \frac{c_5 \left( s \right)}{c_6 \left( s \right)} (e^{\tau_\text{m}}-1) \right]^2 \left( 2 c_1\right)^{-5}\,.
\label{eq:b}
\end{equation}

In Fig.~\ref{f:b_alpha}, we plot the coefficient $b$ as a function of $\alpha=(1-s)/2$. 
The departure of the approximate solutions from exact ones exceeds 5\% for 
$\alpha\lesssim-0.82$.

\begin{table}
\begin{center}
\caption{Amplitude scale corrections for the S2087D VLBA experiment.}
\begin{tabular}{c c c c c}
\hline\noalign{\smallskip}
Antenna &          Band & Epoch &   IF & Correction \\
    (1) &           (2) &   (3) &  (4) &        (5) \\
\hline\noalign{\smallskip}
     BR &  K            &     1 & 1--2 &       0.88 \\
     BR &  K            &     1 & 3--4 &       0.85 \\
     BR &  K            & 2,3,4 & 1--4 &       0.80 \\
     FD &  U            &     1 & 1--4 &       1.09 \\
     FD &  Q            &     1 & 1--4 &       1.15 \\
     KP &  C            &   2,4 & 1--2 &       1.08 \\
     KP &  X            &     1 & 1--2 &       0.90 \\ 
     KP &  X            & 2,3,4 & 1--2 &       0.93 \\
     KP &  K            &   All & 1--4 &       1.10 \\
     LA &  C            &   All & 1--2 &       0.93 \\
     LA &  K            &   All & 1--4 &       0.90 \\
     OV &  X            &     1 & 1--2 &       1.21 \\
     OV &  X            &   3,4 & 1--2 &       1.17 \\
     SC &  U            &     1 & 1--4 &       0.88 \\ 
     SC &  Q            &     2 &  2,4 &       0.80 \\
\hline
\end{tabular}
\end{center}

\medskip
Column designation: (1) antenna name; (2) radio band name; (3)
observation epoch (epochs are labeled as follows: 1 for 2010-05-15, 
2 for 2010-06-25, 3 for 2010-08-01, 4 for 2010-09-09); (4) number of 
the frequency channel (IF); (5) amplitude scale correction coefficient.
\end{table}

\begin{table*}
\caption{Summary of image parameters. Columns are as follows:
(1) epoch of observations,
(2) central observing frequency,
(3) I peak of image,
(4) rms noise level of image,
(5) thermal noise estimate,
(6) bottom I contour level,
(7) dynamic range of image,
(8) total flux density from map,
(9) FWHM major axis of restoring beam,
(10) FWHM minor axis of restoring beam,
(11) position angle of major axis of restoring beam.
}
\begin{tabular}{c r c c c c r c c c r}
\hline\noalign{\smallskip}
Epoch     &  Freq. & $I_\textrm{peak}$ &  $I_\textrm{rms}$ &     Thermal noise & $I_\textrm{base}$ &  DR &$S_\textrm{VLBA}$ & $B_\textrm{maj}$ & $B_\textrm{min}$ & $B_\textrm{PA}$ \\
          &  (GHz) &   (mJy bm$^{-1}$) &   (mJy bm$^{-1}$) &   (mJy bm$^{-1}$) &   (mJy bm$^{-1}$) &     &            (mJy) &            (mas) &            (mas) &           (deg) \\
      (1) &  (2)~~ &               (3) &               (4) &               (5) &               (6) & (7) &              (8) &              (9) &             (10) &            (11) \\
\hline\noalign{\smallskip}
2010-05-15 &  4.608 &   335 &  0.19 &  0.11 &  0.76 &  1756 &   505 &  4.40 &  1.74 &  $-$2.0 \\
2010-06-25 &  4.608 &   408 &  0.15 &  0.11 &  0.61 &  2676 &   569 &  4.97 &  1.87 &  $-$7.6 \\
2010-08-01 &  4.608 &   382 &  0.17 &  0.11 &  0.68 &  2235 &   538 &  4.54 &  1.80 &  $-$3.5 \\
2010-09-09 &  4.608 &   357 &  0.21 &  0.11 &  0.83 &  1723 &   510 &  4.50 &  1.79 &  $-$2.0 \\
2010-05-15 &  5.003 &   350 &  0.18 &  0.15 &  0.71 &  1979 &   519 &  4.14 &  1.65 &  $-$3.3 \\
2010-06-25 &  5.003 &   413 &  0.15 &  0.15 &  0.60 &  2737 &   570 &  4.74 &  1.76 &  $-$8.1 \\
2010-08-01 &  5.003 &   371 &  0.30 &  0.15 &  1.19 &  1243 &   542 &  3.51 &  1.39 &  $-$2.2 \\
2010-09-09 &  5.003 &   359 &  0.17 &  0.15 &  0.68 &  2115 &   514 &  4.21 &  1.67 &  $-$2.2 \\
2010-05-15 &  8.108 &   438 &  0.15 &  0.16 &  0.59 &  2963 &   588 &  2.35 &  0.95 &  $-$0.7 \\
2010-06-25 &  8.108 &   470 &  0.18 &  0.16 &  0.70 &  2677 &   615 &  2.71 &  1.03 &  $-$5.9 \\
2010-08-01 &  8.108 &   417 &  0.17 &  0.16 &  0.69 &  2410 &   566 &  2.46 &  0.97 &  $-$1.7 \\
2010-09-09 &  8.108 &   335 &  0.15 &  0.16 &  0.61 &  2203 &   478 &  2.49 &  0.99 &  $-$1.8 \\
2010-05-15 &  8.429 &   445 &  0.16 &  0.16 &  0.64 &  2778 &   595 &  2.31 &  0.92 &  $-$2.2 \\
2010-06-25 &  8.429 &   477 &  0.14 &  0.16 &  0.56 &  3429 &   624 &  2.68 &  0.99 &  $-$7.5 \\
2010-08-01 &  8.429 &   425 &  0.16 &  0.16 &  0.63 &  2707 &   569 &  2.44 &  0.95 &  $-$3.1 \\
2010-09-09 &  8.429 &   345 &  0.14 &  0.16 &  0.58 &  2396 &   487 &  2.45 &  0.96 &  $-$3.1 \\
2010-05-15 & 15.365 &   557 &  0.20 &  0.18 &  0.68 &  2853 &   697 &  1.58 &  0.49 & $-$11.9 \\
2010-06-25 & 15.365 &   517 &  0.19 &  0.18 &  0.66 &  2735 &   648 &  1.57 &  0.51 & $-$10.2 \\
2010-08-01 & 15.365 &   406 &  0.19 &  0.18 &  0.65 &  2179 &   545 &  1.37 &  0.48 &  $-$6.6 \\
2010-09-09 & 15.365 &   296 &  0.19 &  0.18 &  0.68 &  1523 &   426 &  1.38 &  0.48 &  $-$6.4 \\
2010-05-15 & 23.804 &   557 &  0.26 &  0.21 &  1.02 &  2181 &   685 &  0.96 &  0.29 & $-$12.3 \\
2010-06-25 & 23.804 &   496 &  0.33 &  0.21 &  1.30 &  1521 &   616 &  1.12 &  0.29 & $-$15.4 \\
2010-08-01 & 23.804 &   320 &  0.27 &  0.21 &  1.10 &  1167 &   449 &  0.95 &  0.27 & $-$12.6 \\
2010-09-09 & 23.804 &   250 &  0.26 &  0.21 &  1.06 &   945 &   350 &  1.19 &  0.31 & $-$15.2 \\
2010-05-15 & 43.217 &   511 &  0.40 &  0.32 &  1.41 &  1271 &   631 &  0.49 &  0.17 &  $-$8.5 \\
2010-06-25 & 43.217 &   450 &  0.44 &  0.32 &  1.55 &  1013 &   583 &  0.75 &  0.19 & $-$16.9 \\
2010-08-01 & 43.217 &   442 &  0.92 &  0.45 &  3.23 &   479 &   585 &  0.97 &  0.19 & $-$16.3 \\
2010-09-09 & 43.217 &   346 &  0.96 &  0.45 &  3.36 &   361 &   439 &  0.68 &  0.17 & $-$14.3 \\
\hline
\end{tabular}
\end{table*}

\begin{table*}
\caption{Source models. Columns are as follows:
(1) observation date, 
(2) name of the component,  
(3) flux density of the fitted Gaussian component, 
(4) position offset from the core component,
(5) position angle of the component with respect to the core component, 
(6) FWHM major axis of the fitted Gaussian,
(7) SNR of the fitted Gaussian.
}
\begin{tabular}{c c c c c c r}
\hline\noalign{\smallskip}
      Date & Comp. &     Flux density &        Distance &           P.A. &            Size &  SNR \\
           &       &             (Jy) &           (mas) &          (deg) &           (mas) &      \\
       (1) &   (2) &              (3) &             (4) &            (5) &             (6) &  (7) \\
\hline\noalign{\smallskip}
\multicolumn{7}{c}{4.6 GHz}\\
\hline
2010-05-15 &  Core &  $0.304\pm0.019$ & $0.000\phantom{\,\pm\,0.000}$ &         \ldots & $0.322\pm0.014$ &  535 \\ 
           &    J2 &  $0.127\pm0.012$ & $1.454\pm0.036$ & $113.7\pm 1.4$ & $1.009\pm0.070$ &  207 \\ 
           &    J1 &  $0.068\pm0.015$ & $5.049\pm0.515$ & $100.5\pm 5.8$ & $4.884\pm1.030$ &   23 \\ 
2010-06-25 &  Core &  $0.363\pm0.025$ & $0.000\phantom{\,\pm\,0.000}$ &         \ldots & $0.333\pm0.016$ &  428 \\ 
           &    J2 &  $0.127\pm0.015$ & $1.408\pm0.048$ & $114.2\pm 1.9$ & $1.092\pm0.094$ &  134 \\ 
           &    J1 &  $0.068\pm0.016$ & $5.012\pm0.557$ & $103.0\pm 6.3$ & $4.926\pm1.114$ &   20 \\ 
2010-08-01 &  Core &  $0.350\pm0.020$ & $0.000\phantom{\,\pm\,0.000}$ &         \ldots & $0.356\pm0.014$ &  618 \\ 
           &    J2 &  $0.115\pm0.011$ & $1.394\pm0.040$ & $112.0\pm 1.6$ & $1.039\pm0.078$ &  180 \\ 
           &    J1 &  $0.065\pm0.013$ & $4.997\pm0.461$ & $102.1\pm 5.3$ & $4.706\pm0.922$ &   27 \\ 
2010-09-09 &  Core &  $0.327\pm0.026$ & $0.000\phantom{\,\pm\,0.000}$ &         \ldots & $0.391\pm0.022$ &  315 \\ 
           &    J2 &  $0.109\pm0.014$ & $1.333\pm0.051$ & $115.2\pm 2.1$ & $1.026\pm0.100$ &  104 \\ 
           &    J1 &  $0.066\pm0.014$ & $4.939\pm0.479$ & $101.6\pm 5.5$ & $4.722\pm0.958$ &   25 \\ 
\hline
\multicolumn{7}{c}{5.0 GHz}\\
\hline
2010-05-15 &  Core &  $0.329\pm0.020$ & $0.000\phantom{\,\pm\,0.000}$ &         \ldots & $0.329\pm0.014$ &  547 \\ 
           &    J2 &  $0.117\pm0.012$ & $1.485\pm0.038$ & $113.4\pm 1.4$ & $0.951\pm0.074$ &  166 \\ 
           &    J1 &  $0.067\pm0.016$ & $5.057\pm0.572$ & $101.9\pm 6.5$ & $4.918\pm1.144$ &   19 \\ 
2010-06-25 &  Core &  $0.382\pm0.027$ & $0.000\phantom{\,\pm\,0.000}$ &         \ldots & $0.326\pm0.016$ &  387 \\ 
           &    J2 &  $0.114\pm0.014$ & $1.425\pm0.046$ & $112.4\pm 1.8$ & $0.997\pm0.090$ &  125 \\ 
           &    J1 &  $0.067\pm0.014$ & $4.967\pm0.508$ & $102.7\pm 5.8$ & $5.053\pm1.016$ &   26 \\ 
2010-08-01 &  Core &  $0.360\pm0.021$ & $0.000\phantom{\,\pm\,0.000}$ &         \ldots & $0.349\pm0.014$ &  560 \\ 
           &    J2 &  $0.117\pm0.012$ & $1.376\pm0.043$ & $113.3\pm 1.7$ & $1.113\pm0.084$ &  173 \\ 
           &    J1 &  $0.059\pm0.014$ & $5.088\pm0.499$ & $103.1\pm 5.6$ & $4.484\pm0.998$ &   21 \\ 
2010-09-09 &  Core &  $0.331\pm0.021$ & $0.000\phantom{\,\pm\,0.000}$ &         \ldots & $0.382\pm0.018$ &  474 \\ 
           &    J2 &  $0.115\pm0.013$ & $1.295\pm0.047$ & $115.3\pm 2.0$ & $1.107\pm0.092$ &  145 \\ 
           &    J1 &  $0.060\pm0.013$ & $5.033\pm0.452$ & $101.2\pm 5.1$ & $4.413\pm0.904$ &   25 \\ 
\hline
\multicolumn{7}{c}{8.1 GHz}\\
\hline
2010-05-15 &  Core &  $0.449\pm0.032$ & $0.000\phantom{\,\pm\,0.000}$ &         \ldots & $0.250\pm0.012$ &  376 \\ 
           &    J2 &  $0.098\pm0.016$ & $1.491\pm0.073$ & $112.4\pm 2.8$ & $1.096\pm0.146$ &   57 \\ 
           &    J1 &  $0.044\pm0.019$ & $5.387\pm0.931$ & $101.0\pm 9.8$ & $4.357\pm1.862$ &    6 \\ 
2010-06-25 &  Core &  $0.469\pm0.031$ & $0.000\phantom{\,\pm\,0.000}$ &         \ldots & $0.246\pm0.012$ &  438 \\ 
           &    J2 &  $0.106\pm0.015$ & $1.374\pm0.076$ & $113.0\pm 3.2$ & $1.269\pm0.152$ &   72 \\ 
           &    J1 &  $0.043\pm0.017$ & $5.323\pm0.871$ & $102.0\pm 9.3$ & $4.383\pm1.742$ &    7 \\ 
2010-08-01 &  Core &  $0.420\pm0.030$ & $0.000\phantom{\,\pm\,0.000}$ &         \ldots & $0.269\pm0.014$ &  393 \\ 
           &    J2 &  $0.103\pm0.016$ & $1.303\pm0.087$ & $113.2\pm 3.8$ & $1.313\pm0.174$ &   58 \\ 
           &    J1 &  $0.045\pm0.018$ & $5.244\pm0.893$ & $101.3\pm 9.7$ & $4.421\pm1.786$ &    7 \\ 
2010-09-09 &  Core &  $0.335\pm0.023$ & $0.000\phantom{\,\pm\,0.000}$ &         \ldots & $0.311\pm0.016$ &  397 \\ 
           &    J2 &  $0.101\pm0.014$ & $1.216\pm0.076$ & $114.8\pm 3.6$ & $1.284\pm0.152$ &   72 \\ 
           &    J1 &  $0.043\pm0.017$ & $5.199\pm0.916$ & $101.5\pm10.0$ & $4.500\pm1.832$ &    7 \\ 
\hline
\multicolumn{7}{c}{8.4 GHz}\\
\hline
2010-05-15 &  Core &  $0.455\pm0.034$ & $0.000\phantom{\,\pm\,0.000}$ &         \ldots & $0.233\pm0.012$ &  359 \\ 
           &    J2 &  $0.098\pm0.017$ & $1.495\pm0.082$ & $112.7\pm 3.1$ & $1.137\pm0.164$ &   50 \\ 
           &    J1 &  $0.044\pm0.021$ & $5.352\pm1.036$ & $103.6\pm11.0$ & $4.382\pm2.072$ &    5 \\ 
2010-06-25 &  Core &  $0.480\pm0.030$ & $0.000\phantom{\,\pm\,0.000}$ &         \ldots & $0.236\pm0.010$ &  508 \\ 
           &    J2 &  $0.099\pm0.015$ & $1.410\pm0.074$ & $112.8\pm 3.0$ & $1.217\pm0.148$ &   69 \\ 
           &    J1 &  $0.044\pm0.017$ & $5.318\pm0.830$ & $103.1\pm 8.9$ & $4.350\pm1.660$ &    8 \\ 
2010-08-01 &  Core &  $0.429\pm0.030$ & $0.000\phantom{\,\pm\,0.000}$ &         \ldots & $0.255\pm0.012$ &  411 \\ 
           &    J2 &  $0.100\pm0.015$ & $1.313\pm0.085$ & $112.6\pm 3.7$ & $1.304\pm0.170$ &   59 \\ 
           &    J1 &  $0.043\pm0.017$ & $5.261\pm0.853$ & $102.6\pm 9.2$ & $4.453\pm1.706$ &    8 \\ 
2010-09-09 &  Core &  $0.343\pm0.026$ & $0.000\phantom{\,\pm\,0.000}$ &         \ldots & $0.289\pm0.016$ &  336 \\ 
           &    J2 &  $0.102\pm0.015$ & $1.187\pm0.080$ & $114.3\pm 3.9$ & $1.291\pm0.160$ &   66 \\ 
           &    J1 &  $0.045\pm0.018$ & $5.144\pm0.878$ & $102.7\pm 9.7$ & $4.551\pm1.756$ &    8 \\ 
\hline
\end{tabular}
\end{table*}

\begin{table*}
\contcaption{}
\begin{tabular}{c c c c c c r}
\hline\noalign{\smallskip}
      Date & Comp. &     Flux density &        Distance &           P.A. &            Size &  SNR \\
           &       &             (Jy) &           (mas) &          (deg) &           (mas) &      \\
       (1) &   (2) &              (3) &             (4) &            (5) &             (6) &  (7) \\
\hline\noalign{\smallskip}
\multicolumn{7}{c}{15.4 GHz}\\
\hline
2009-12-26 &  Core &  $0.383\pm0.021$ & $0.000\phantom{\,\pm\,0.000}$ &         \ldots & $0.120\pm0.004$ &  680 \\ 
           &    J3 &  $0.051\pm0.008$ & $0.421\pm0.029$ & $114.6\pm 3.9$ & $0.447\pm0.058$ &   59 \\ 
           &    J2 &  $0.061\pm0.008$ & $1.715\pm0.021$ & $111.0\pm 0.7$ & $0.430\pm0.042$ &  102 \\ 
           &    J1 &  $0.030\pm0.018$ & $5.341\pm1.434$ & $101.5\pm15.0$ & $4.726\pm2.868$ &    4 \\
2010-05-15 &  Core &  $0.584\pm0.039$ & $0.000\phantom{\,\pm\,0.000}$ &         \ldots & $0.156\pm0.008$ &  448 \\
           &    J3 &  $0.037\pm0.010$ & $0.619\pm0.035$ & $115.0\pm 3.2$ & $0.334\pm0.070$ &   24 \\
           &    J2 &  $0.049\pm0.010$ & $1.740\pm0.043$ & $111.2\pm 1.4$ & $0.530\pm0.086$ &   40 \\
           &    J1 &  $0.031\pm0.024$ & $5.299\pm1.679$ & $ 98.0\pm17.6$ & $4.379\pm3.358$ &    2 \\
2010-06-19 &  Core &  $0.577\pm0.027$ & $0.000\phantom{\,\pm\,0.000}$ &         \ldots & $0.130\pm0.006$ &  922 \\
           &    J3 &  $0.059\pm0.009$ & $0.520\pm0.034$ & $119.2\pm 3.7$ & $0.358\pm0.068$ &   86 \\
           &    J2 &  $0.046\pm0.007$ & $1.719\pm0.037$ & $111.3\pm 1.2$ & $0.625\pm0.074$ &   79 \\
           &    J1 &  $0.034\pm0.017$ & $5.218\pm1.150$ & $100.3\pm12.4$ & $4.714\pm2.300$ &    5 \\
2010-06-25 &  Core &  $0.534\pm0.029$ & $0.000\phantom{\,\pm\,0.000}$ &         \ldots & $0.154\pm0.006$ &  650 \\
           &    J3 &  $0.045\pm0.009$ & $0.589\pm0.028$ & $118.4\pm 2.7$ & $0.377\pm0.056$ &   46 \\
           &    J2 &  $0.044\pm0.007$ & $1.727\pm0.042$ & $110.8\pm 1.4$ & $0.615\pm0.084$ &   55 \\
           &    J1 &  $0.029\pm0.021$ & $5.411\pm1.624$ & $102.0\pm16.7$ & $4.606\pm3.248$ &    3 \\
2010-08-01 &  Core &  $0.439\pm0.031$ & $0.000\phantom{\,\pm\,0.000}$ &         \ldots & $0.200\pm0.010$ &  380 \\
           &    J3 &  $0.041\pm0.010$ & $0.641\pm0.032$ & $118.4\pm 2.9$ & $0.358\pm0.064$ &   32 \\
           &    J2 &  $0.043\pm0.008$ & $1.716\pm0.057$ & $112.9\pm 1.9$ & $0.705\pm0.114$ &   40 \\
           &    J1 &  $0.027\pm0.020$ & $5.335\pm1.530$ & $100.1\pm16.0$ & $4.187\pm3.060$ &    3 \\
2010-09-09 &  Core &  $0.298\pm0.025$ & $0.000\phantom{\,\pm\,0.000}$ &         \ldots & $0.190\pm0.012$ &  274 \\
           &    J3 &  $0.055\pm0.012$ & $0.348\pm0.035$ & $100.4\pm 5.6$ & $0.404\pm0.068$ &   37 \\
           &    J2 &  $0.057\pm0.017$ & $1.474\pm0.164$ & $110.6\pm 6.3$ & $1.198\pm0.328$ &   14 \\
           &    J1 &  $0.022\pm0.017$ & $5.771\pm1.463$ & $ 99.1\pm14.2$ & $3.910\pm2.926$ &    2 \\
2010-12-24 &  Core &  $0.306\pm0.019$ & $0.000\phantom{\,\pm\,0.000}$ &         \ldots & $0.000\pm0.010$ &  526 \\ 
           &    J3 &  $0.060\pm0.009$ & $0.442\pm0.031$ & $ 98.9\pm 4.0$ & $0.171\pm0.062$ &   88 \\ 
           &    J2 &  $0.046\pm0.009$ & $1.411\pm0.077$ & $110.5\pm 3.1$ & $0.790\pm0.154$ &   36 \\ 
           &    J1 &  $0.023\pm0.022$ & $4.943\pm2.275$ & $ 95.5\pm24.7$ & $4.094\pm4.550$ &    2 \\ 
2011-09-12 &  Core &  $0.644\pm0.039$ & $0.000\phantom{\,\pm\,0.000}$ &         \ldots & $0.000\pm0.008$ &  559 \\ 
           &    J3 &  $0.071\pm0.014$ & $0.269\pm0.073$ & $103.3\pm15.2$ & $0.756\pm0.146$ &   32 \\ 
           &    J2 &  $0.088\pm0.015$ & $1.611\pm0.044$ & $110.7\pm 1.6$ & $0.494\pm0.088$ &   54 \\ 
           &    J1 &  $0.029\pm0.032$ & $5.275\pm2.814$ & $101.1\pm28.1$ & $4.401\pm5.628$ &    1 \\ 
2012-05-24 &  Core &  $1.029\pm0.034$ & $0.000\phantom{\,\pm\,0.000}$ &         \ldots & $0.090\pm0.002$ & 1785 \\ 
           &    J3 &  $0.053\pm0.008$ & $0.506\pm0.032$ & $ 94.3\pm 3.6$ & $0.286\pm0.064$ &   79 \\ 
           &    J2 &  $0.089\pm0.011$ & $1.612\pm0.031$ & $113.0\pm 1.1$ & $0.609\pm0.062$ &  100 \\ 
           &    J1 &  $0.027\pm0.016$ & $5.417\pm1.109$ & $101.0\pm11.6$ & $3.927\pm2.218$ &    4 \\ 
2012-07-12 &  Core &  $0.716\pm0.028$ & $0.000\phantom{\,\pm\,0.000}$ &         \ldots & $0.140\pm0.004$ & 1273 \\ 
           &    J3 &  $0.066\pm0.009$ & $0.538\pm0.026$ & $ 98.9\pm 2.8$ & $0.328\pm0.052$ &   94 \\ 
           &    J2 &  $0.082\pm0.012$ & $1.561\pm0.034$ & $114.0\pm 1.2$ & $0.585\pm0.068$ &   74 \\ 
           &    J1 &  $0.040\pm0.032$ & $4.142\pm2.124$ & $ 97.8\pm27.1$ & $5.239\pm4.248$ &    2 \\ 
2012-12-10 &  Core &  $0.857\pm0.027$ & $0.000\phantom{\,\pm\,0.000}$ &         \ldots & $0.062\pm0.002$ & 1970 \\ 
           &    J3 &  $0.049\pm0.006$ & $0.644\pm0.027$ & $ 98.6\pm 2.4$ & $0.447\pm0.054$ &   96 \\ 
           &    J2 &  $0.071\pm0.009$ & $1.569\pm0.027$ & $113.0\pm 1.0$ & $0.534\pm0.054$ &   99 \\ 
           &    J1 &  $0.035\pm0.020$ & $4.964\pm1.369$ & $101.0\pm15.4$ & $4.717\pm2.738$ &    4 \\ 
2016-09-17 &  Core &  $0.470\pm0.031$ & $0.000\phantom{\,\pm\,0.000}$ &         \ldots & $0.061\pm0.010$ &  472 \\ 
           &    J3 &  $0.103\pm0.016$ & $0.401\pm0.036$ & $105.2\pm 5.1$ & $0.347\pm0.072$ &   72 \\ 
           &    J2 &  $0.072\pm0.016$ & $1.695\pm0.091$ & $109.0\pm 3.1$ & $0.838\pm0.182$ &   29 \\ 
           &    J1 &  $0.030\pm0.021$ & $5.485\pm1.554$ & $105.8\pm15.8$ & $4.362\pm3.108$ &    3 \\
\hline
\end{tabular}
\end{table*}

\begin{table*}
\contcaption{}
\begin{tabular}{c c c c c c r}
\hline\noalign{\smallskip}
      Date & Comp. &     Flux density &        Distance &           P.A. &            Size &  SNR \\
           &       &             (Jy) &           (mas) &          (deg) &           (mas) &      \\
       (1) &   (2) &              (3) &             (4) &            (5) &             (6) &  (7) \\
\hline\noalign{\smallskip}
\multicolumn{7}{c}{23.8 GHz}\\
\hline
2010-05-15 &  Core &  $0.498\pm0.026$ & $0.000\phantom{\,\pm\,0.000}$ &         \ldots & $0.044\pm0.002$ &  754 \\
           &    J4 &  $0.112\pm0.012$ & $0.147\pm0.003$ & $ 92.9\pm 1.2$ & $0.000\pm0.006$ &  165 \\
           &    J3 &  $0.026\pm0.006$ & $0.600\pm0.039$ & $114.5\pm 3.7$ & $0.389\pm0.078$ &   26 \\
           &    J2 &  $0.033\pm0.007$ & $1.756\pm0.052$ & $109.9\pm 1.7$ & $0.542\pm0.104$ &   28 \\
2010-06-25 &  Core &  $0.436\pm0.032$ & $0.000\phantom{\,\pm\,0.000}$ &         \ldots & $0.009\pm0.002$ &  376 \\
           &    J4 &  $0.109\pm0.017$ & $0.164\pm0.006$ & $101.2\pm 2.1$ & $0.000\pm0.012$ &   86 \\
           &    J3 &  $0.027\pm0.008$ & $0.661\pm0.038$ & $116.4\pm 3.3$ & $0.320\pm0.076$ &   19 \\
           &    J2 &  $0.031\pm0.009$ & $1.744\pm0.081$ & $110.0\pm 2.7$ & $0.652\pm0.162$ &   17 \\
2010-08-01 &  Core &  $0.246\pm0.022$ & $0.000\phantom{\,\pm\,0.000}$ &         \ldots & $0.013\pm0.004$ &  258 \\
           &    J4 &  $0.140\pm0.017$ & $0.163\pm0.006$ & $104.4\pm 2.1$ & $0.131\pm0.012$ &  135 \\
           &    J3 &  $0.019\pm0.005$ & $0.788\pm0.024$ & $123.0\pm 1.7$ & $0.228\pm0.048$ &   23 \\
           &       &  $0.007\pm0.003$ & $1.389\pm0.042$ & $116.5\pm 1.7$ & $0.200\pm0.084$ &   11 \\
           &    J2 &  $0.019\pm0.004$ & $1.858\pm0.041$ & $112.6\pm 1.3$ & $0.420\pm0.082$ &   27 \\
2010-09-09 &  Core &  $0.185\pm0.023$ & $0.000\phantom{\,\pm\,0.000}$ &         \ldots & $0.027\pm0.008$ &  135 \\
           &    J4 &  $0.111\pm0.018$ & $0.175\pm0.012$ & $107.6\pm 3.6$ & $0.192\pm0.022$ &   75 \\
           &    J3 &  $0.021\pm0.007$ & $0.763\pm0.031$ & $118.3\pm 2.3$ & $0.257\pm0.062$ &   19 \\
           &       &  $0.008\pm0.003$ & $1.330\pm0.040$ & $113.2\pm 1.7$ & $0.200\pm0.080$ &   15 \\
           &    J2 &  $0.017\pm0.005$ & $1.844\pm0.062$ & $111.5\pm 1.9$ & $0.511\pm0.124$ &   18 \\
\hline
\multicolumn{7}{c}{43.2 GHz}\\
\hline
2010-05-15 &  Core &  $0.487\pm0.025$ & $0.000\phantom{\,\pm\,0.000}$ &         \ldots & $0.026\pm0.000$ &  740 \\ 
           &    J4 &  $0.112\pm0.014$ & $0.129\pm0.005$ & $ 91.4\pm 2.2$ & $0.100\pm0.010$ &  124 \\ 
           &    J3 &  $0.015\pm0.005$ & $0.667\pm0.062$ & $120.0\pm 5.3$ & $0.379\pm0.124$ &   10 \\ 
           &    J2 &  $0.019\pm0.007$ & $1.806\pm0.111$ & $105.3\pm 3.5$ & $0.606\pm0.222$ &    8 \\ 
2010-06-25 &  Core &  $0.412\pm0.025$ & $0.000\phantom{\,\pm\,0.000}$ &         \ldots & $0.028\pm0.002$ &  559 \\ 
           &    J4 &  $0.133\pm0.014$ & $0.134\pm0.006$ & $ 87.9\pm 2.6$ & $0.145\pm0.012$ &  156 \\ 
           &    J2 &  $0.019\pm0.007$ & $1.813\pm0.097$ & $110.8\pm 3.1$ & $0.595\pm0.194$ &   10 \\ 
           &    J3 &  $0.018\pm0.005$ & $0.686\pm0.029$ & $116.3\pm 2.4$ & $0.247\pm0.058$ &   20 \\ 
2010-08-01 &  Core &  $0.444\pm0.030$ & $0.000\phantom{\,\pm\,0.000}$ &         \ldots & $0.051\pm0.002$ &  431 \\ 
           &    J4 &  $0.103\pm0.016$ & $0.183\pm0.005$ & $ 94.9\pm 1.6$ & $0.005\pm0.010$ &   88 \\ 
           &    J3 &  $0.022\pm0.008$ & $0.761\pm0.035$ & $ 95.5\pm 2.6$ & $0.225\pm0.070$ &   14 \\ 
           &    J2 &  $0.018\pm0.006$ & $1.980\pm0.030$ & $115.7\pm 0.9$ & $0.238\pm0.060$ &   17 \\ 
2010-09-09 &  Core &  $0.241\pm0.019$ & $0.000\phantom{\,\pm\,0.000}$ &         \ldots & $0.011\pm0.002$ &  309 \\ 
           &    J4 &  $0.067\pm0.010$ & $0.197\pm0.011$ & $104.7\pm 3.2$ & $0.190\pm0.022$ &   77 \\ 
           &    J3 &  $0.014\pm0.005$ & $0.840\pm0.020$ & $118.3\pm 1.4$ & $0.128\pm0.040$ &   18 \\ 
           &    J2 &  $0.013\pm0.005$ & $1.765\pm0.030$ & $107.1\pm 1.0$ & $0.153\pm0.060$ &   12 \\ 
\hline
\end{tabular}
\end{table*}

\begin{table*}
\centering
\caption{Image shift, core offset shift, and core shift vectors measured for the frequency pairs $\nu_1$ and $\nu_2$.}
\begin{tabular}{c r c c c r r r c}
\hline\noalign{\smallskip}
Epoch & $\nu_1$ ~$\nu_2$ & $\Delta r_{12}$ & $r_1-r_2$ & $\Delta r_{\text{core},\nu_1\nu_2}$ & P.A.$_{\Delta r_{12}}$ &    P.A.$_{r_1-r_2}$ &    P.A.$_{\Delta r_{\text{core},\nu_1\nu_2}}$ & $\lambda_2-\lambda_1$ \\
      &            (GHz) &           (mas) &     (mas) &                               (mas) &                (deg) &             (deg) &                                       (deg) &                  (cm) \\
  (1) &            (2)~~ &             (3) &       (4) &                                 (5) &                  (6) &               (7) &                                         (8) &                   (9) \\
\hline\noalign{\smallskip}
2010-05-15 &  43.2 \,~4.6 &  0.794 &  0.189 &  0.615 &   101 &     117 &     96 &  5.816 \\
2010-05-15 &  43.2 \,~5.0 &  0.671 &  0.139 &  0.539 &   100 &     117 &     96 &  5.302 \\
2010-05-15 &  43.2 \,~8.1 &  0.242 &  0.010 &  0.244 &    97 &  $-$159 &     95 &  3.006 \\
2010-05-15 &  43.2 \,~8.4 &  0.285 &  0.014 &  0.278 &   108 &     171 &    106 &  2.865 \\
2010-05-15 &  43.2   15.4 &  0.090 &  0.013 &  0.088 &    90 &     168 &     82 &  1.258 \\
2010-05-15 &  43.2   23.8 &  0.030 &  0.012 &  0.037 &    90 &  $-$152 &     73 &  0.566 \\
2010-05-15 &  23.8 \,~4.6 &  0.713 &  0.189 &  0.527 &   105 &     113 &    102 &  5.250 \\
2010-05-15 &  23.8 \,~5.0 &  0.626 &  0.140 &  0.487 &   107 &     112 &    105 &  4.736 \\
2010-05-15 &  23.8 \,~8.1 &  0.190 &  0.003 &  0.188 &   108 &      53 &    109 &  2.440 \\
2010-05-15 &  23.8 \,~8.4 &  0.228 &  0.009 &  0.220 &   113 &     111 &    113 &  2.299 \\
2010-05-15 &  23.8   15.4 &  0.085 &  0.009 &  0.078 &   135 &     102 &    138 &  0.692 \\
2010-05-15 &  15.4 \,~4.6 &  0.560 &  0.181 &  0.382 &   106 &     114 &    102 &  4.558 \\
2010-05-15 &  15.4 \,~5.0 &  0.532 &  0.131 &  0.401 &   106 &     112 &    104 &  4.044 \\
2010-05-15 &  15.4 \,~8.1 &  0.153 &  0.007 &  0.160 &   101 &   $-$59 &    102 &  1.748 \\
2010-05-15 &  15.4 \,~8.4 &  0.153 &  0.001 &  0.153 &   101 &  $-$161 &    101 &  1.607 \\
2010-05-15 &   8.4 \,~4.6 &  0.474 &  0.181 &  0.295 &   108 &     113 &    105 &  2.951 \\
2010-05-15 &   8.4 \,~5.0 &  0.313 &  0.131 &  0.183 &   107 &     112 &    103 &  2.437 \\
2010-05-15 &   8.1 \,~4.6 &  0.342 &  0.188 &  0.159 &   105 &     114 &     95 &  2.810 \\
2010-05-15 &   8.1 \,~5.0 &  0.371 &  0.138 &  0.235 &   104 &     113 &     99 &  2.296 \\

2010-06-25 &  43.2 \,~4.6 &  0.886 &  0.169 &  0.720 &   114 &     123 &    112 &  5.816 \\
2010-06-25 &  43.2 \,~5.0 &  0.780 &  0.143 &  0.637 &   113 &     118 &    111 &  5.302 \\
2010-06-25 &  43.2 \,~8.1 &  0.295 &  0.041 &  0.331 &   114 &   $-$35 &    118 &  3.006 \\
2010-06-25 &  43.2 \,~8.4 &  0.309 &  0.033 &  0.339 &   119 &   $-$38 &    121 &  2.865 \\
2010-06-25 &  43.2   15.4 &  0.095 &  0.014 &  0.106 &   108 &   $-$32 &    113 &  1.258 \\
2010-06-25 &  43.2   23.8 &  0.090 &  0.026 &  0.108 &    90 &   $-$39 &    101 &  0.566 \\
2010-06-25 &  23.8 \,~4.6 &  0.793 &  0.194 &  0.600 &   119 &     126 &    117 &  5.250 \\
2010-06-25 &  23.8 \,~5.0 &  0.741 &  0.168 &  0.573 &   122 &     122 &    122 &  4.736 \\
2010-06-25 &  23.8 \,~8.1 &  0.242 &  0.015 &  0.255 &   120 &   $-$27 &    122 &  2.440 \\
2010-06-25 &  23.8 \,~8.4 &  0.216 &  0.007 &  0.223 &   124 &   $-$34 &    124 &  2.299 \\
2010-06-25 &  23.8   15.4 &  0.067 &  0.012 &  0.056 &   117 &     132 &    113 &  0.692 \\
2010-06-25 &  15.4 \,~4.6 &  0.698 &  0.182 &  0.519 &   115 &     125 &    112 &  4.558 \\
2010-06-25 &  15.4 \,~5.0 &  0.564 &  0.156 &  0.409 &   115 &     121 &    113 &  4.044 \\
2010-06-25 &  15.4 \,~8.1 &  0.162 &  0.027 &  0.185 &   112 &   $-$36 &    116 &  1.748 \\
2010-06-25 &  15.4 \,~8.4 &  0.134 &  0.019 &  0.152 &   117 &   $-$43 &    119 &  1.607 \\
2010-06-25 &   8.4 \,~4.6 &  0.484 &  0.201 &  0.285 &   120 &     126 &    115 &  2.951 \\
2010-06-25 &   8.4 \,~5.0 &  0.443 &  0.174 &  0.270 &   118 &     123 &    116 &  2.437 \\
2010-06-25 &   8.1 \,~4.6 &  0.417 &  0.208 &  0.212 &   120 &     128 &    113 &  2.810 \\
2010-06-25 &   8.1 \,~5.0 &  0.417 &  0.181 &  0.236 &   120 &     124 &    117 &  2.296 \\
2010-08-01 &  23.8 \,~4.6 &  0.698 &  0.173 &  0.525 &   115 &     113 &    116 &  5.250 \\
2010-08-01 &  23.8 \,~5.0 &  0.591 &  0.138 &  0.454 &   114 &     107 &    116 &  4.736 \\
2010-08-01 &  23.8   15.4 &  0.085 &  0.009 &  0.080 &   135 &      80 &    140 &  0.692 \\
2010-08-01 &  23.8 \,~8.1 &  0.242 &  0.035 &  0.207 &   120 &     127 &    118 &  2.440 \\
2010-08-01 &  23.8 \,~8.4 &  0.201 &  0.013 &  0.188 &   117 &     102 &    118 &  2.299 \\
2010-08-01 &  15.4 \,~4.6 &  0.671 &  0.166 &  0.505 &   117 &     114 &    117 &  4.558 \\
2010-08-01 &  15.4 \,~5.0 &  0.552 &  0.130 &  0.421 &   112 &     109 &    113 &  4.044 \\
2010-08-01 &  15.4 \,~8.1 &  0.175 &  0.030 &  0.147 &   121 &     140 &    117 &  1.748 \\
2010-08-01 &  15.4 \,~8.4 &  0.134 &  0.006 &  0.128 &   117 &     135 &    116 &  1.607 \\
2010-08-01 &   8.4 \,~4.6 &  0.457 &  0.160 &  0.297 &   113 &     114 &    113 &  2.951 \\
2010-08-01 &   8.4 \,~5.0 &  0.379 &  0.125 &  0.254 &   108 &     108 &    109 &  2.437 \\
2010-08-01 &   8.1 \,~4.6 &  0.418 &  0.139 &  0.279 &   111 &     109 &    112 &  2.810 \\
2010-08-01 &   8.1 \,~5.0 &  0.313 &  0.106 &  0.208 &   107 &     101 &    110 &  2.296 \\
2010-09-09 &  43.2   23.8 &  0.090 &  0.044 &  0.058 &   179 &     146 & $-$156 &  0.566 \\
2010-09-09 &  23.8 \,~4.6 &  0.631 &  0.073 &  0.642 &   115 &  $-$149 &    109 &  5.250 \\
2010-09-09 &  23.8 \,~5.0 &  0.525 &  0.131 &  0.578 &   121 &  $-$131 &    109 &  4.736 \\
2010-09-09 &  23.8 \,~8.1 &  0.268 &  0.032 &  0.263 &   117 &  $-$167 &    110 &  2.440 \\
2010-09-09 &  23.8 \,~8.4 &  0.256 &  0.014 &  0.270 &   111 &   $-$50 &    112 &  2.299 \\
2010-09-09 &  23.8   15.4 &  0.060 &  0.033 &  0.085 &    90 &   $-$40 &    107 &  0.692 \\
2010-09-09 &  15.4 \,~4.6 &  0.457 &  0.089 &  0.446 &   113 &  $-$170 &    102 &  4.558 \\
2010-09-09 &  15.4 \,~5.0 &  0.433 &  0.136 &  0.456 &   124 &  $-$145 &    106 &  4.044 \\
2010-09-09 &  15.4 \,~8.1 &  0.201 &  0.058 &  0.170 &   117 &     166 &    101 &  1.748 \\
2010-09-09 &  15.4 \,~8.4 &  0.124 &  0.019 &  0.111 &   104 &     147 &     97 &  1.607 \\
2010-09-09 &   8.4 \,~4.6 &  0.258 &  0.076 &  0.249 &   126 &  $-$160 &    108 &  2.951 \\
2010-09-09 &   8.4 \,~5.0 &  0.162 &  0.130 &  0.158 &   158 &  $-$137 &    110 &  2.437 \\
2010-09-09 &   8.1 \,~4.6 &  0.295 &  0.044 &  0.313 &   114 &  $-$136 &    106 &  2.810 \\
2010-09-09 &   8.1 \,~5.0 &  0.234 &  0.107 &  0.287 &   130 &  $-$121 &    109 &  2.296 \\
\hline
\end{tabular}
\end{table*}


\bsp	
\label{lastpage}
\end{document}